\documentclass[osajnl,twocolumn,showpacs,superscriptaddress,10pt]{revtex4-1}

\usepackage{amsmath,amssymb,graphicx}

\usepackage{color}
\usepackage{aas_macros}

\newcommand{\degs}{^\circ}
\newcommand{\vect}[1]{\boldsymbol{#1}}
\newcommand{\ex}{\vect{\hat{e}_x}}
\newcommand{\ey}{\vect{\hat{e}_y}}
\newcommand{\ez}{\vect{\hat{e}_z}}
\newcommand{\er}{\vect{\hat{e}_r}}
\newcommand{\eq}{\vect{\hat{e}_\theta}}
\newcommand{\ep}{\vect{\hat{e}_\phi}}
\newcommand{\ev}{\vect{\hat{e}_v}}
\newcommand{\eh}{\vect{\hat{e}_h}}
\newcommand{\evb}{\vect{\hat{e}_{\bar{v}}}}
\newcommand{\ehb}{\vect{\hat{e}_{\bar{h}}}}
\newcommand{\eperp}{\vect{\hat{e}_\perp}}
\newcommand{\eparal}{\vect{\hat{e}_\parallel}}

\begin{document}

\title{Angular and Polarization Response of Multimode Sensors \\ with Resistive-Grid Absorbers}

\author{Akito Kusaka}\email{Corresponding author: akusaka@princeton.edu}
\affiliation{Department of Physics, Princeton University, Princeton, NJ, USA}

\author{Edward J. Wollack}
\author{Thomas R. Stevenson}
\affiliation{NASA Goddard Space Flight Center, Greenbelt, MD, USA}

\begin{abstract}
 High sensitivity receiver systems with near ideal polarization
 sensitivity are highly desirable for development of millimeter and
 sub-millimeter radio astronomy.  Multimoded bolometers provide a unique
 solution to achieve such sensitivity, for which hundreds of single-mode
 sensors would otherwise be required. The primary concern in employing
 such multimoded sensors for polarimetery is the control of the
 polarization systematics. In this paper, we examine the angular- and
 polarization- dependent absorption pattern of a thin resistive grid or
 membrane, which models an absorber used for a multimoded bolometer. The
 result shows that a freestanding thin resistive absorber with a
 surface resistivity of $\eta/2$, where $\eta$ is the impedance of free space, attains a beam pattern with equal $E$- and
 $H$-plane responses, leading to zero cross polarization.
 For a resistive-grid absorber, the condition is met when
 a pair of grids is positioned orthogonal to each other and both 
 have a resistivity of $\eta/2$.
 When a
 reflective backshort termination is employed to improve absorption
 efficiency, the cross-polar level can be suppressed below $-30$\,dB if
 acceptance angle of the sensor is limited to $\lesssim
 60\degs$.
 The small cross-polar systematics have even-parity patterns
 and
 do not contaminate the measurements of odd-parity
 polarization patterns, for which many of recent instruments for cosmic
 microwave  background are designed.
 Underlying symmetry
 that suppresses these cross-polar systematics is discussed in
 detail. The estimates and formalism provided in this paper offer key
 tools in the design consideration of the instruments using the
 multimoded polarimeters.
\end{abstract}

\maketitle

\section{Introduction}
Present astronomical instrumentation applications in the
millimeter and sub-millimeter desire photon backgrounded limited
sensitivities.  There are two possible basic approaches to further
improve the sensitivity by increasing the number of detected spatial
modes received by an imaging system.  The first is to build an array
consisting of numerous single-mode sensors with high optical
efficiency.  The second, a multi-mode sensor, detects many spatial modes
on a single sensor (see, e.g., Ref.~\cite{1981IJQE...17..407R})
with well-defined angular and polarization
characteristics.
In astronomical observations, for example, both single-mode sensors and
multimoded sensors (see, e.g., Refs.~\cite{Benford2008,Lawrence2008})
have found wide use for radiometry and photometry.

\begin{figure*}[tbp]
 \begin{center}
  \includegraphics[width=0.9\textwidth]{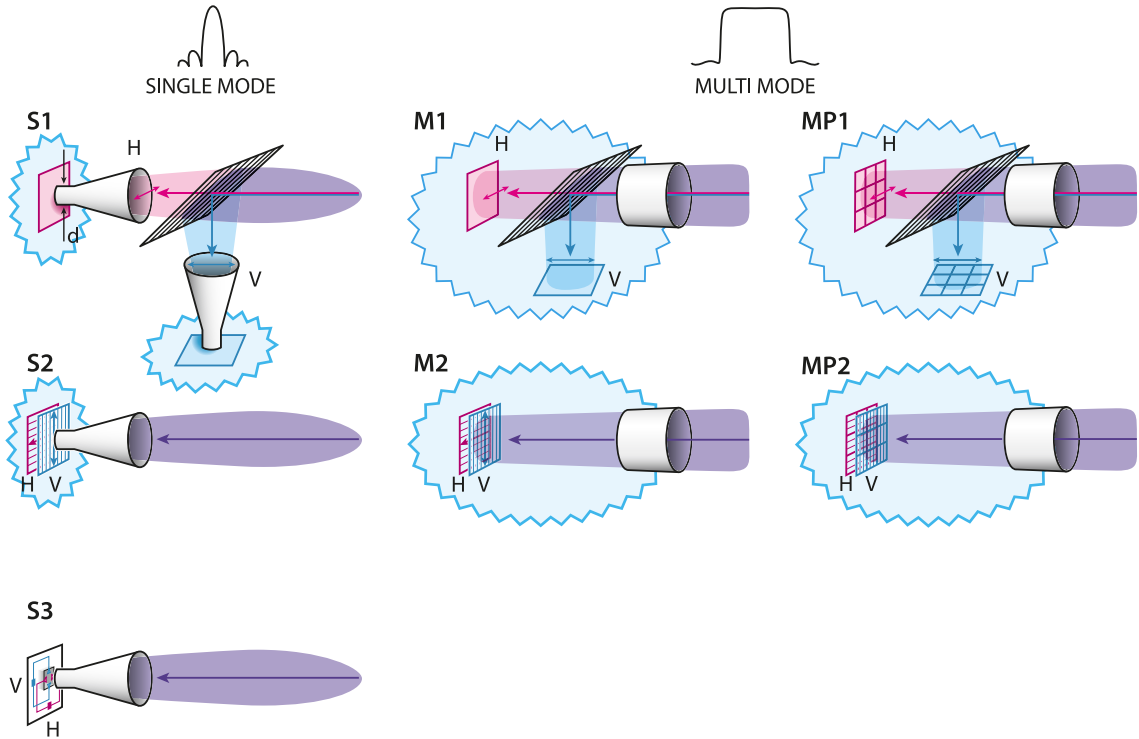}
  \caption{ \label{fig:filter_techniques} 
  Various modal filtering techniques for polarization sensitive detector
  systems.
  Throughout this paper, a single-mode dual-polarization sensor
  denotes a system
  that consists of a pair of single-mode detectors that are sensitive to two
  orthogonal polarizations.
  S1: a single-mode dual-polarization sensor comprising two intrinsically-multimoded
  polarization-insensitive sensors.
  Here, single-mode limit is set by the feed-exit coupling port
  (symbolically shown as the diameter ``$d$'' of the port appropriately set compared to
  the wavelength) and 
  polarization separation is achieved via quasi-optical grid diplexer.
  The blue structures surrounding each sensor correspond to cold
  baffling.
  S2: a single-mode dual-polarization sensor comprising intrinsically-multimoded
  polarization-sensitive bolometers (PSBs),
  where the single-mode limit is set in the same way as S1 and the
  polarization separation is achieved via the patterned absorber
  structure of the bolometer.
  S3: a single-mode dual-polarization sensor where the radiation input is split
  into two polarization components by
  a single-mode orthomode transducer (OMT) and subsequently absorbed by
  a pair of detectors.  The feedhorn here merely provides a controlled
  coupling
  of the single-mode
  set by the detector structure
  to a single-mode in free-space;
  it can be replaced by an immersion lens or a phased-array antenna
  structure.
  M1: a multimoded polarimetor comprising polarization-insensitive
  multimoded sensors.  The number-of-modes limit is set by
  the baffling limiting the solid angle seen from the sensor and the
  polarization separation
  is achieved via quasi-optical grid diplexer (the same as S1).
  MP1: a polarimetor array consisting of
  polarization-insensitive segmented sensors filling the space
  and a quasi-optical grid diplexer.
  The spatial-mode limit and polarization
  separation are achieved in the same way as M1.
  Each pixel
  can be either single- or multi-moded
  depending on the size of
  the pixels relative to the wavelength and the solid-angle limit set by
  the baffling.
  M2: a multimoded polarimetor comprising a pair of multimoded
  polarization-sensitive bolometers,
  where the number-of-modes limit is set in the same way as M1 and the
  polarization separation is achieved via the patterned absorber
  structure of the bolometer.
  MP2: a polarimetor array consisting of polarization-sensitive
  filled-array sensors.  The spatial-mode limit and polarization
  separation are achieved in the same way as M2.
  }
 \end{center}
\end{figure*}

Various techniques can be used to specify the modes coupled to a sensor
and the resulting system architectures can be categorized by their modal
filtering techniques (Fig.~\ref{fig:filter_techniques}).
In the context of polarimetry at millimeter wavelengths,
where a significant use is for measuring
the cosmic microwave background (CMB),
the primary
focus of the recent developments has been directed at large arrays of
single-mode detectors with feed-coupled waveguide polarization
diplexers~\cite{2009AIPC.1185..494E,
2010SPIE.7741E..51N, CLASS.SPIE.2012,
2009AIPC.1185..511M}
or planar-antenna coupled
structures~\cite{2010SPIE.7741E..40O,
2010SPIE.7741E..50S,
2011AA...536A...1P,
POLAR.SPIE.2012,
2010SPIE.7741E..39A,
2008SPIE.7010E..79C}
 which can
be photolithographically produced in large numbers.
Here, polarization
diplexing is achieved on the detector chip for the horizontal and
vertical single mode detector channels
(Fig.~\ref{fig:filter_techniques}, S3). Prior to these
developments, dual-mode waveguide-based orthomode transducer (OMT) structures followed by
single-mode detectors~\cite{2003ApJS..145..413J,
2013ApJ...768....9B,
Clover2009}
 were used to form a
polarimeter following the traditions of microwave design. Alternatively,
single-mode dual-polarization
 sensors were realized by combining intrinsically multimoded
polarization-sensitive bolometers (PSBs) with external modal filtering
structures to define the angular acceptance
(Fig.~\ref{fig:filter_techniques},
S2)~\cite{2007A&A...470..771J,2010ApJ...711.1141T,2008ApOpt..47.5996H,2009ApJ...692.1221H}.
Multimoded sensors for imaging have also achieved polarization
sensitivity through a wire-grid analyzer~\cite{1997PASP..109..307S,2005ASPC..343...69B,2006SPIE.6275E..48L,2010SPIE.7741E..49B}
(Fig.~\ref{fig:filter_techniques}, M1 and MP1). The analyzer
grid architecture has also been employed in conjunction with a feed
coupled array (e.g., 
Ref.~\cite{2004SPIE.5543..320O}) 
 to cleanly
provide polarization sensitivity (Fig.~\ref{fig:filter_techniques}, S1).

A multimoded bolometric sensor that is  intrinsically polarization
selective, as opposed to those using the wire-grid analyzer,
belongs to another class of sensors and opens up a new phase space for
polarization sensitive instruments in millimeter
 wavelength (Fig.~\ref{fig:filter_techniques}, M2 and MP2).  Such a
bolometer would employ a thin resistive grid as a polarization selective
absorber.  An M2 implementation is the sensor
developed for
PIXIE satellite mission proposal aimed at CMB polarization and
frequency spectrum measurements~\cite{2011JCAP...07..025K}.
This sensor employs a pair of orthogonally positioned resistive grids,
each of which is separately read out,
attaining simultaneous sensitivity to two linear
polarizations.
Other applications
exploiting the unique features of this sensor are
also proposed~\cite{doi:10.1117/12.926271}.
An MP2 implementation, an intrinsically polarization-selective sensor
based on a pair of
filled arrays, was explored in Ref.~\cite{2006SPIE.6275E..27W} where each
detector absorber $\sim \lambda^2$
on a side was realized from a patterned thin metal film on a Si membrane
and spaced by $\sim$ microns (Fig.~\ref{fig:filter_techniques}, MP2).
These types of sensors (M2 and MP2), similarly to M1 and MP1, require
cold baffling to control the detector power loading, but share
an improved mapping speed advantage compared to the single-mode
sensors~\cite{2002ApOpt..41.6543G}.

The focus of this paper is to provide an estimate of the angular- and
polarization-dependent response pattern of the thin resistive absorber,
and to show that bolometers corresponding to the M2 type in Fig.~\ref{fig:filter_techniques} can achieve low levels of
polarization beam systematics.
Further, the residual non-zero systematics are shown to have even-parity
patterns.
A potential use for these sensors is to
probe the signature of the inflation in the early universe through the
odd-parity, or so-called $B$-mode, patterns in the CMB
polarization~\cite{1997PhRvL..78.2054S,1997PhRvL..78.2058K},
which is not contaminated by the even-parity beam
systematics~\cite{2003PhRvD..67d3004H,2008PhRvD..77h3003S}.
Implications presented in this paper are general and applicable to a wide
range of devices using thin absorbing grids or membranes,
even though our work is motivated by the specific implementation
of the device mentioned above~\cite{2011JCAP...07..025K,doi:10.1117/12.926271}.
We choose to analytically study this system -- for numerical
investigations the approach and method of Ref.~\cite{Thomas:13} could
be adopted.
The results we present would serve as a key tool in the design consideration for
this class of sensors and the instruments employing them.
Our analysis gives attention to the response pattern at large off-axis
incident angles.  The use of the large off-axis angles
makes stray-light control easier and
 is often desirable
in maximizing the number of modes.
The number of modes ($N_\mathrm{modes}$)
is related to the absorption area ($A$), the
solid angle ($\Omega$) and the wavelength ($\lambda$) by
 $A \Omega = N_\mathrm{modes} \, \lambda^2$, and larger $\Omega$ 
allows $N_\mathrm{modes}$ to be maximized
 while restraining the increase of the
physical detector size $A$.
A combination of a fast final-focus lens and a multimoded
polarimetor (Fig.~\ref{fig:filter_techniques}, M2) would provide such
a large solid
angle, for example.  We note that so-called filled-array sensors
(Fig.~\ref{fig:filter_techniques}, MP1 or MP2) are often deployed in a
similar configuration using a
final-focus lens.  For this reason, the results presented in this paper
offer potentially useful considerations for the filled-array sensors
with thin resistive-membrane absorbers ~\cite{2004SPIE.5492.1064H,2006SPIE.6275E..42D,2008SPIE.7020E..44T,2006SPIE.6275E..44S}
 even though they may be
used as dual polarization photometers in
the single to several mode limit.

There are a few assumptions in our analysis regarding the configuration
of the device of
interest.  First, we assume
the absorbing grid or membrane is resistive, as opposed to reactive, and
the current in the direction normal to the absorber plane can be ignored.
These approximations are valid if the absorber's physical thickness is
small compared to the penetration depth of the resistive coating
({\it electrically thin}, hereafter).
Secondly, when the absorber is a pair of orthogonal resistive grids, which are
sensitive to orthogonal polarizations, we assume
that the pitch of each grid and 
the
distance between the two grids are small compared to the wavelength,
and thus they are effectively resistive sheets on the same plane.
It can be shown that there is little near-field modal coupling between the crossed grids in this regime.
In Appendix~\ref{app:detector_geometry_assumption}, we discuss the
conditions on the physical dimension of the detector
such that the above stated approximations are valid.  
For example, the device proposed in Ref.~\cite{2011JCAP...07..025K}
satisfies such conditions.
Under these assumptions, we can treat the
grid pair and the membrane equivalently, except the grid pair may
have different resistivity in the two orthogonal directions.

We start by setting up a formalism to evaluate the polarization
 systematics in Sec.~\ref{sec:contrast_to_singlemode}.
We review the standard
measure of the polarization systematics such as cross polarization
in a single-mode system, and extend them
to a multimoded sensor that comprises an electrically thin resistive absorber.
Section~\ref{sec:s_matrix_vs_far_field_functions} provides a rigorous
 foundation of this extension through an $S$-matrix
 formalism.
 Notably, this discussion 
 provides ground that generalizes the results
 presented in Sec.~\ref{sec:elemag_calc} to an arbitrary incident mode,
 whereas the derivation
 in Sec.~\ref{sec:elemag_calc} is based on plane-wave incident modes.
In Sec.~\ref{sec:elemag_calc}, we discuss the result of electromagnetic
 calculations
for the response pattern of an electrically thin resistive grid or
 membrane absorber
 sheet with
an infinitely large area,
with and without reflective backshort termination.
We then briefly discuss the systematics due to diffraction of a
finite-sized absorber in Sec.~\ref{sec:diffraction_consideration}.  

\section{Polarization Systematics} \label{sec:contrast_to_singlemode}
\begin{figure}[tbp]
 \includegraphics[width=0.47\textwidth]{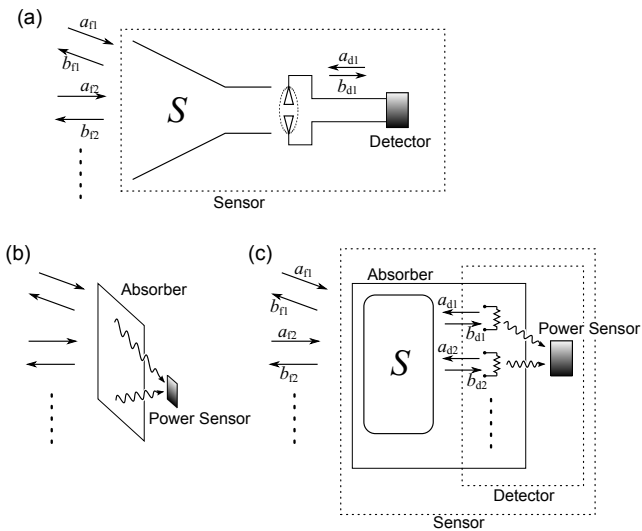}
 \caption{\label{fig:sensor_detector_clarification} 
 Schematic figures showing our definition of the {\it detector}
 and the {\it sensor},
 and the mode couplings to them,
 for the examples of (a) a single-mode feedhorn-coupled bolometer
 and (b) a multimoded bolometer using a sheet absorber.
 For the former, a feedhorn serves as a coupling to the electromagnetic
 waves in free space, and a bolometer serves as a {\it detector} that converts the
 electromagnetic waves to an electric signal.
 Here, $a_{f1}, a_{f2}, \cdots$
 and $b_{f1}, b_{f2}, \cdots$
 are the amplitudes of incoming and radiating plane waves,
 respectively,
 and $a_{d1}$ and $b_{d1}$ are the amplitude of the
 radiation from the
 detector and that absorbed by the detector, respectively.
 An $S$ matrix characterizes
 the coupling among these modes provided by the feedhorn.
 For a multimoded bolometer in (b), on the other hand, the absorber
 both provides the coupling to the free-space--propagating waves and
 converts electromagnetic waves to thermal signals.
 The thermal signals then
 propagate to
 the power sensor (e.g., thermistor or TES) and are read out as an electric signal.
 Panel (c) schematically models (b) and clarifies the two roles the
 absorber plays and the fact that multiple modes couples to a power sensor,
 where $S$ in the figure implies the coupling among the modes.
 Here, the {\it detector} corresponds to the combination of the dissipating
 part in the  absorber and the power sensor.
 }
\end{figure}

In the context of polarimetry, our focus in evaluating the goodness of a
response pattern of a sensor is for control of polarization systematics.
In particular, low cross polarization and small differential
polarization response are important for high polarization efficiency and low
spurious polarization~\cite{2003PhRvD..67d3004H,2008PhRvD..77h3003S}, respectively.
In this section, we describe the formalism to evaluate the ability of
the sensor to reject these artifacts in the angular response and retain
the polarization purity of the incident radiation.
Throughout this section, the word {\it sensor}
represents the entire polarimeter including the coupling to the
plane wave propagating in free space,
while the word {\it detector} denotes
 a part
that converts an incoming
electromagnetic wave to an electric signal that can be
read out~(Fig.~\ref{fig:sensor_detector_clarification}).

It is convenient to discuss the polarization systematics of a multimoded
system
in contrast to that of a single-mode system,
which has already been discussed in detail in
literature.
We will adopt Ludwig's third definition~\cite{1973ITAP...21..116L},
in which the electric field directions of the two
linear polarization bases are
\begin{equation}
 \begin{split}
  \ev & = \eq \cos \phi - \ep \sin \phi \:, \quad \mathrm{and}
  \\
  \eh & = \eq \sin \phi + \ep \cos \phi \:,
 \end{split}
\end{equation}
where
$\eq$ and $\ep$ are unit vectors in spherical coordinates 
(see Appendix~\ref{sec:vect_def}).
As a convention, we define the $z$ axis as the on-axis direction of the optics
and take the standard definition of a spherical coordinate system
specified by
$(r, \theta, \phi)$ where $\theta$ is the angle from the $z$ axis.
The vertical (horizontal) polarization direction $\ev$ ($\eh$)
asymptotes to the $x$ ($y$) axis direction for $\theta \rightarrow 0$.

It is customary to characterize the beam 
of a single-mode sensor through the radiation process.
Although relevant for a receiver
instrument is the reception process, time-reversal symmetry allows us to
na\"{i}vely relate the
characteristics in the radiation process to those in the reception
process.
For a single-mode sensor nominally sensitive to vertical [horizontal]
polarization,
the radiation field pattern
is described using co- and cross-polar far-field functions,
$G^{v[h]}_{CO}(\theta,\phi)$ and $G^{v[h]}_{XP}(\theta,\phi)$, as
\begin{equation}
 \begin{split}
  \vect{E}^{v[h]}&(r,\theta,\phi) 
  \\
  \propto & \frac{e^{-i k r}}{r}
  \left\{ G^{v[h]}_{CO}(\theta,\phi) \, \vect{\hat{e}_{v[h]}}
  + G^{v[h]}_{XP}(\theta,\phi) \, \vect{\hat{e}_{h[v]}}
  \right\}
  \:,
 \end{split}
 \label{equ:radiation_field_by_Gco_Gxp}
\end{equation}
where $k$ is the wave number ($k=2\pi / \lambda$)
and the superscript $v$ or $h$ (vertical or horizontal) denotes the
nominal polarization of
the detector.  
In practical applications, a polarimeter is often a dual-polarization
sensor equipped with two detectors nominally measuring two orthogonal
polarizations.  We refer to such a dual polarization sensor as a
single-mode system in the limit each polarization sensitive detector
couples a single mode and the polarization isolation can be treated as a
subdominant perturbation to the sensor response.

The beam systematics are quantified using the far-field
functions.  We define the cross-polar level $XP(\theta)$
as cross-polar radiation power:
\begin{equation}
 XP(\theta)
  \equiv \max_\phi \left|
   \frac{G^v_{XP}(\theta,\phi)}{G^v_{CO}(0,0)}
   \right|^2
  \:,
  \label{equ:cross_polar_level}
\end{equation}
where we indicate the worst case in $\phi$ by $\max_\phi$.
Note that we do not assume normalization of the far-field functions 
to the
on-axis beam and thus the denominator in
Eq.~(\ref{equ:cross_polar_level}).
The lack of normalization,
as opposed to what is often standard,
is our intentional choice since it is more appropriate when describing a
multimoded system as we see later.
The differential response is defined
as the difference of co-polar radiation power
between the
two
detectors:
\begin{equation}
 DR(\theta)
  \equiv \frac{1}{2}
  \max_\phi
  \left|
   \left|
   \frac{G_{CO}^v(\theta,\phi)}
   {G_{CO}^v(0,0)}\right|^2 
   -
   \left|
   \frac{G_{CO}^h(\theta,\phi)}
  {G_{CO}^h(0,0)}\right|^2
  \right|
 \:,
    \label{equ:differential_response}
\end{equation}
where we again take the worst case in $\phi$.
Zero systematics, i.e., $XP=0$ and $DR=0$, are attained
when the far-field complex gain response satisfies
\begin{equation}
 \begin{split}
  G_{CO}^v(\theta,\phi) & = G_{CO}^h(\theta,\phi) = G(\theta) \:,
  \\
  G_{XP}^v(\theta,\phi) & = G_{XP}^h(\theta,\phi) = 0 \:.
 \end{split}
 \label{equ:zero_systematic_condition}
\end{equation}

We briefly comment on some general properties of the far-field
functions and the systematics measures.
These properties apply to a scalar feed
as well as a multimoded 
absorber sheet with isotropic resistivity,
which we will show later.
If the sensor of interest is symmetric under a $90\degs$
rotation about the $z$ axis,
the far-field functions satisfy the following relation:
\begin{equation}
 \begin{split}
 G_{CO}^v(\theta,\phi)
  & = G_{CO}^h(\theta,\phi + \pi/2)
  \\
   G_{XP}^v(\theta,\phi)
  & = - G_{XP}^h(\theta,\phi + \pi/2) \:,
 \end{split}
\end{equation}
since the $90\degs$ rotational operation acts on the
elements in Eq.~(\ref{equ:radiation_field_by_Gco_Gxp})
as
$\left(\vect{E_v},\vect{E_h}\right)
  \rightarrow \left(-\vect{E_h},\vect{E_v}\right)$
and
$\left(\ev,\eh\right) \rightarrow \left(-\eh,\ev\right)$.
This justifies the definition of Eq.~(\ref{equ:cross_polar_level}),
which may otherwise appear different for $p=h$.
When the coupling to free space possesses continuous rotational symmetry
about the $z$ axis,
the $\phi$ to achieve
$\max_\phi$ in Eqs.~(\ref{equ:cross_polar_level}) and
(\ref{equ:differential_response})
corresponds to $\phi=\pi/4$ and $\phi=0$,
respectively~\cite{kildal2000foundations}. 
$DR(\theta)$ corresponds to the difference of the co-polar beams for
the E- and H-planes when this condition is achieved.
It is known that a necessary and sufficient condition
for a scalar feed
to attain zero systematics [satisfy
Eq.~(\ref{equ:zero_systematic_condition})]
is to posses symmetric $E$- and $H$-plane responses,
or $G^v_{CO}(\theta,0) = G^v_{CO}(\theta,\pi/2)$~\cite{1973ITAP...21..116L,kildal2000foundations,stutzman1997atd,2010ITAP...58.1383Z}.
The orders of magnitudes of $XP$ and $DR$ are typically related as
$DR \sim \sqrt{ XP }$ since
$\left|G_{CO}^v - G_{CO}^h \right| \sim \left| G_{XP} \right|$.

The radiation field pattern is widely used to describe the beam pattern
and its systematics for a single-mode system.  In contrast, in this
paper we mainly
discuss reception processes in deriving the beam pattern of a multimoded
sensor.
This is because a multimoded beam pattern can be represented as a
combination of multiple radiation field patterns and the relative
excitation strengths among the radiation modes, and the latter is more
naturally derived through reception processes.
We assume a multimoded sensor with an electrically thin resistive
absorber~(Fig.~\ref{fig:sensor_detector_clarification})
lying on the $x$-$y$ plane.
In the 
 reception process, an incident plane wave induces
absorber-surface current $j_x$ ($j_y$) in the $x$ ($y$) direction.
We  relate 
the induced current, $j_p$, and 
 the amplitude of an incident
plane wave, $a_p(\theta,\phi)$, with an incident angle $(\theta,\phi)$ and a polarization
$p$ 
by co- and cross-polar coupling coefficients,
$\widetilde{G}^p_{CO}(\theta,\phi)$ and
$\widetilde{G}^p_{XP}(\theta,\phi)$:
\begin{equation}
 \left(
  \begin{array}{c}
   \rho_x j_x \\
   \rho_y j_y
  \end{array}
 \right)
 =
  \left(
   \begin{array}{cc}
    \widetilde{G}_{CO}^{v}(\theta,\phi) & \widetilde{G}_{XP}^{v}(\theta,\phi) \\
    \widetilde{G}_{XP}^{h}(\theta,\phi) & \widetilde{G}_{CO}^{h}(\theta,\phi)
   \end{array}
  \right)
 \left(
  \begin{array}{c}
   a_v(\theta,\phi) \\
   a_h(\theta,\phi)
  \end{array}
 \right) \:,
 \label{equ:G__ampl_vs_current}
\end{equation}
where $\rho_{x(y)}$ is the absorber surface resistivity in the $x(y)$
direction.
One would build a dual-polarization sensor by using a stacked pair
of resistive grids as the absorber sheet, where each of the grids
is aligned in
the $x$ or $y$ direction,
and by coupling the thermal signals from $j_x$ and $j_y$ 
to different power sensors~\cite{2011JCAP...07..025K}.

Since the system of interest is multimoded,
it is implicit in Eq.~(\ref{equ:G__ampl_vs_current}) that the surface
current has multiple excitation modes
for each of $j_x$ and $j_y$.
The modes may be expressed in terms of two-dimensional Fourier modes of
the 
surface current, and the modal content of the induced current is dependent on
$(\theta,\phi,p)$, the
incident angle and polarization of the plane wave.
We will clarify the details of the modal content in the next section.
The coefficients $\widetilde{G}^p_{CO[XP]}(\theta,\phi)$
represent the couplings between the current and the plane waves
 regardless of the modal content.
When the system of interest is single-moded, 
the modal content does not depend on $(\theta,\phi,p)$ and 
$\widetilde{G}^p_{CO[XP]}(\theta,\phi)$
is equated to the far-field functions
$G^p_{CO[XP]}(\theta,\phi)$ 
through the symmetry between the radiation and reception processes.
This is not the case for a multimoded system;
e.g., the $\widetilde{G}^p_{CO[XP]}(\theta,\phi)$ 
do not describe a radiation field pattern as in
Eq.~(\ref{equ:radiation_field_by_Gco_Gxp}).
As discussed in the next section, however,
we can regard $\widetilde{G}^p_{CO[XP]}(\theta,\phi)$
equivalently to the far-field functions
for an infinitely large, electrically thin resistive absorber sheet.
For example, we show below that one can substitute them
into Eqs.~(\ref{equ:cross_polar_level}) and
(\ref{equ:differential_response})
in place of $G^p_{CO[XP]}(\theta,\phi)$
to evaluate the beam systematics.

\section{Beam Characterization Using the Coupling Coefficients $\vect{\widetilde{G}}$}
\label{sec:s_matrix_vs_far_field_functions}
In this section,
we derive the relation between the coupling coefficients
$\widetilde{G}^p_{CO[XP]}(\theta,\phi)$
and the beam characteristics, in particular the beam systematics,
for an infinitely large, electrically thin resistive absorber sheet.
We first introduce a formalism using a scattering matrix,
and relate it to the far-field functions of single-mode and
multimoded systems.
This allows us to generalize the expressions of
the beam pattern,
and thus 
Eqs.~(\ref{equ:cross_polar_level}) and
(\ref{equ:differential_response}), for a multimoded system.
We then show in the limit of infinitely large, electrically thin
resistive
absorbers an incident plane wave with a specific $(\theta,\phi,p)$ couples
to only a single surface-current mode.
This simplifies the generalized expressions
back to those for a single-mode system,
except the far-field functions 
$G^p_{CO[XP]}(\theta,\phi)$
are replaced by the coupling coefficients
$\widetilde{G}^p_{CO[XP]}(\theta,\phi)$.
This formalism also shows that our result is applicable to an arbitrary
incident mode, even though the definition of
Eq.~(\ref{equ:G__ampl_vs_current}) and the
drivation in Sec.~\ref{sec:elemag_calc} assume plane-wave incident
modes.  This generalization is of importance since spherical
waves are the incident
modes when a detector is placed at the focus of a telescope.
However, the derivation in Sec.~\ref{sec:elemag_calc} does not require
this groundwork.

\subsection{$\boldsymbol{S}$-Matrix Formalism}
A general response of a sensor can be described using a scattering
 matrix, or $S$ matrix, relating the incoming and outgoing modes
(see, e.g., Ref.~\cite{2003ApOpt..42.4989Z} for a
 formal discussion).
Incoming modes are either plane waves from the sky propagating in free
 space, whose  amplitudes
are $a_{f1}, a_{f2}, \cdots$,
or radiation from the detector, whose amplitudes are
$a_{d1}, a_{d2}, \cdots$.
Outgoing modes,
which are the time reversal of the incoming modes,
 are either plane waves reflected toward the sky
with amplitudes $b_{f1}, b_{f2}, \cdots$,
or the modes absorbed by the detector
with amplitudes $b_{d1}, b_{d2}, \cdots$~(Fig.~\ref{fig:sensor_detector_clarification}).
The amplitudes are proportional to the electric field strength 
and follow a Gaussian random distribution for radiation from
a thermal source (e.g., the CMB).
An $S$ matrix describes
the coupling of the incoming and outgoing modes
provided by an optical coupling element (e.g., a feedhorn)
and relates the amplitudes as
\begin{equation}
 b_i(\nu) = \sum_j S_{ij}(\nu) \, a_j(\nu) \:,
  \label{equ:mode_propagation}
\end{equation}
where the indices $i$ and $j$ runs over both
$f1, f2, \cdots$
and $d1, d2, \cdots$.
For simplicity, we omit the dependence on frequency $\nu$ hereafter. 

A single-mode sensor measures a single mode among the outgoing
modes, which we label as $d1$.
The $S$-matrix elements $S_{d1\,j}$ relate the incoming plane-wave
amplitudes $a_{j} (j=f1, f2, \cdots)$ and the measured
amplitude
 $b_{d1}$:
\begin{equation}
 b_{d1} = \sum_{j=f1, f2, \cdots}
  S_{d1\,j} \, a_{j} \:.
  \label{equ:single_moded_response_ampl}
\end{equation}
The detected power $P$ is calculated as
\begin{equation}
 \begin{split}
 P & = \bigl< \bigl|b_{d1}\bigr|^2 \bigr>
  =  \bigl<   \bigl| \sum_{j=f1, f2, \cdots}
  S_{d1\,j} \, a_{j} \bigr|^2 \bigr>
  \\
  & = \sum_{j=f1, f2, \cdots}
  \bigl| S_{d1\,j} \bigr|^2 \bigl< \bigl| a_j \bigr|^2 \bigr>
  \:,
 \end{split}
 \label{equ:single_moded_response}
\end{equation}
where the last equality holds when incoming plane wave amplitudes
are uncorrelated to each other.
We omit the coefficient,
with the dimension of admittance,
relating $P$ and $\bigl|b_{d1}\bigr|^2$.
Since the index $j$ runs over the plane wave modes
with different incident angles
and polarization, $\bigl| S_{d1\,j} \bigr|^2$
corresponds to
the angular and polarization dependent antenna power pattern of the
sensor in the reception process.

It is customary to define the far-field functions $G_{CO}$ and $G_{XP}$
in the context of radiation patterns
and thus they are related to the $S$-matrix elements for the radiation process,
$S_{i\,d1}$.
The system of interest is usually symmetric under
time-reversal operation~\footnote{Rare exceptions include
magnetic devices like circulators.}
and thus the $S$ matrix is Hermitian, $S_{ij} = S_{ji}^*$.
The Hermitian property of the $S$ matrix allows us
to equivalently relate the
far-field functions to the $S$-matrix elements for 
the reception process, $S_{d1\,j}$, 
which are conceptually more straightforward to evaluate for a multimoded
system.

A natural extension of Eqs.~(\ref{equ:single_moded_response_ampl})
and (\ref{equ:single_moded_response})
describes 
the response pattern of a multimoded sensor.  The formalism
differs from a single-mode
system in that there are more than
one mode coupled to the detector
(Fig.~\ref{fig:sensor_detector_clarification}).
The detector-absorbed amplitudes $b_i$ are:
\begin{equation}
 b_i
  =
 \sum_{j = f1, f2, \cdots} 
  S_{ij} \, a_j
  \quad (i = d1, d2, \cdots) \:,
 \label{equ:mm_ampl_s_matrix}
\end{equation}
and the detected power $P$ is:
\begin{equation}
 \label{equ:mm_power_s_matrix}
 \begin{split}
  P & = \Bigl< \sum_{i = d1, d2, \cdots} \left|  b_i \right|^2 \Bigr> 
  =  \Bigl< \sum_{i = d1, d2, \cdots}
  \Bigl| \sum_{j = f1, f2, \cdots} 
  S_{ij} \, a_j
  \Bigr|^2 \Bigr> 
  \\
  &
  = \sum_{i = d1, d2, \cdots}  \Biggl|
  \sum_{j = f1, f2, \cdots} 
	     S_{ij} 
	    \Biggr|^2
  \bigl< \bigl| a_j \bigr|^2 \bigr>  \:,
 \end{split} 
\end{equation}
where, again, the last equality holds when the incoming plane wave
amplitudes are uncorrelated.

The absorbed-mode basis set, with amplitude
$b_i$ ($i=d1, d2, \cdots$), is chosen such that 
each of the modes couples to an eigenmode of the dissipation
process
in the device~(Fig.~\ref{fig:sensor_detector_clarification}c).
This corresponds to the order of operations
in Eq.~(\ref{equ:mm_power_s_matrix})
that the square of
the absolute value, $\left|  b_i \right|^2$,
is taken first and then summed over the index $i$.
In the subsection~\ref{subsec:simpl_infty_large_absorber}
we show that for an infinitely-large, electrically-thin resistive
absorber, two-dimensional Fourier modes of the surface currents serves
as such eigenmodes.

\subsection{Far-Field Functions of Single-Mode Sensors \label{subsec:far_field_in_single_mode}}
Here, we relate the $S$ matrix to the far-field functions
of a single-mode dual-polarization sensor.
For convenience,
we label the two detectors and indices of the detector-coupled modes
by their nominal polarizations, $v$ (vertical) and $h$ (horizontal),
instead of $d1$ and $d2$.
Equation~(\ref{equ:single_moded_response_ampl}) is now rewritten as a
pair of equations:
\begin{equation}
 \left(
  \begin{array}{c}
   b_{v} \\
   b_{h}
  \end{array}
 \right)
 =
 \sum_{j=f1, f2, \cdots}
 \left(
  \begin{array}{c}
   S_{vj}  \\
   S_{hj} 
  \end{array}
 \right) a_{j} \: .
  \label{equ:s_matrix_vs_ampl_vh}
\end{equation}
Since the index $j$ runs over free-space--propagating plane waves,
we can choose the basis set such that the label $j$
corresponds to a set of
 an incident
angle $(\theta, \phi)$ and a linear polarization $p$ $(=v,h)$.
Both $S$ matrix and plane-wave amplitudes are relabeled as
$S_{ij} \rightarrow S_{ip}(\theta,\phi)$ and
$a_j \rightarrow a_p(\theta,\phi)$, respectively.
Equation~(\ref{equ:s_matrix_vs_ampl_vh}) can be conveniently written in
terms of a $2\times 2$ matrix
\begin{equation}
 \vect{b} = 
  \int \! \mathrm{d}\Omega \,
  \vect{S}(\theta,\phi) \vect{a}(\theta,\phi) \:,
  \label{equ:b_vs_S_vs_a_in_single_mode}
\end{equation}
with
\begin{equation}
 \begin{split}
  \vect{a}(\theta,\phi)
  \equiv &
  \left(
  \begin{array}{c}
   a_v(\theta,\phi)\\
   a_h(\theta,\phi)
  \end{array}
  \right) \:,
  \quad 
  \vect{b}
  \equiv
  \left(
  \begin{array}{c}
   b_v\\
   b_h
  \end{array}
  \right) \:, 
  \\
  \vect{S}(\theta,\phi)
  & \equiv
  \left(
  \begin{array}{cc}
   S_{vv}(\theta,\phi) & S_{vh}(\theta,\phi) \\
   S_{hv}(\theta,\phi) & S_{hh}(\theta,\phi)
  \end{array}
  \right) \:,
 \end{split}
 \label{equ:single_mode_a_b_S_definition}
\end{equation}
and
$\mathrm{d}\Omega \equiv \sin \theta \, \mathrm{d}\phi \,
\mathrm{d}\theta$.
Note that the matrix $\vect{S}(\theta,\phi)$ does not have to be
symmetric since it is an off-diagonal block of the entire $S$ matrix,
which is clear in Eq.~(\ref{equ:s_matrix_vs_ampl_vh}).

We then equate the matrix $\vect{S}(\theta,\phi)$ to the far-field
functions:
\begin{equation}
  \left(
  \begin{array}{cc}
   S_{vv}(\theta,\phi) & S_{vh}(\theta,\phi) \\
   S_{hv}(\theta,\phi) & S_{hh}(\theta,\phi)
  \end{array}
  \right)
  =
  \left(
   \begin{array}{cc}
    G_{CO}^v(\theta,\phi) & G_{XP}^v(\theta,\phi) \\
    G_{XP}^h(\theta,\phi) & G_{CO}^h(\theta,\phi)
   \end{array}
  \right) \:.
  \label{equ:single_mode_S_vs_G}
\end{equation}
As noted earlier,
$G_{CO}^p(\theta,\phi)$ and $G_{XP}^p(\theta,\phi)$
slightly deviate from the standard definition in that they are not
normalized to
$G^p_{CO}(0,0)$.  Instead,
their normalization carries additional information about absorption
efficiency.

Now we can rephrase the definitions 
of the beam systematics
in the context of the reception process.
The cross-polar level,
 $XP(\theta)$ defined by Eq.~(\ref{equ:cross_polar_level}),
is the response in power to a cross-polar incident wave
normalized to the response to a co-polar on-axis incident wave.
The differential response,
$DR(\theta)$ defined by Eq.~(\ref{equ:differential_response}),
is the difference of the power response
between the
two detectors to 
their co-polar incident waves
 with the same incident angles.
This language, in contrast to that of the radiation process,
is directly applicable to a multimoded system, too.

\subsection{Far-Field Functions of Multimoded Sensors}
We relate the $S$ matrix to the far-field functions
and define the systematics measures
for a multimoded dual-polarization sensor
by extending those for a single-mode sensor.
We relabel the index of the detector-coupled modes,
$d1, d2, \cdots$,
by a pair $(p,i')$, where $p$ ($= v,h$) labels each of the two detectors
and the index $i'$ runs over the modes coupled to each detector.
Hereafter, we use symbols with primes for variables indexing the detector-coupled modes for clearity.

As is done for the derivation of Eq.~(\ref{equ:b_vs_S_vs_a_in_single_mode}),
we rewrite Eq.~(\ref{equ:mm_ampl_s_matrix}) as:
\begin{equation}
 \vect{b}^{i'} = 
  \int_\Omega \! \mathrm{d}\Omega \,
  \vect{S}^{i'}(\theta,\phi) \vect{a}(\theta,\phi)
  \label{equ:s_matrix_vs_ampl_multimode_2x2matrix}
  \:,
\end{equation}
where
$\vect{b}^{i'}$ and $\vect{S}^{i'}(\theta,\phi)$ are defined
similarly to Eq.~(\ref{equ:single_mode_a_b_S_definition}) but with an
additional index $i'$.
We then define the far-field functions for a multimoded system
as a natural extension of Eq.~(\ref{equ:single_mode_S_vs_G}):
\begin{equation}
\renewcommand{\arraystretch}{1.5}
  \left(
  \begin{array}{cc}
   S^{i'}_{vv}(\theta,\phi) & S^{i'}_{vh}(\theta,\phi) \\
   S^{i'}_{hv}(\theta,\phi) & S^{i'}_{hh}(\theta,\phi)
  \end{array}
  \right)
  =
  \left(
   \begin{array}{cc}
    G_{CO}^{v,{i'}}(\theta,\phi) & G_{XP}^{v,{i'}}(\theta,\phi) \\
    G_{XP}^{h,{i'}}(\theta,\phi) & G_{CO}^{h,{i'}}(\theta,\phi)
   \end{array}
  \right) \:.
  \label{equ:multi_mode_S_vs_G}
\renewcommand{\arraystretch}{1}
\end{equation}
We note again that the normalizations of the far-field functions here
are related to the absorption efficiency;
they are not normalized to on-axis response.
This is more important here than it is for a single-mode system
since the relative normalization among $G_{CO[XP]}^{p,i'}(\theta,\phi)$
for different $i'$ tells us the relative excitation strengths among
different detector-coupled modes.  

We adopt the definitions of the levels of cross polarization and
differential response 
rephrased in the context of reception process
using response power:
\begin{equation}
 XP(\theta)
  \equiv \max_\phi \frac{
     \sum_{i'} \left| G^{v,i'}_{XP}(\theta,\phi)\right|^2}{ \sum_{i'} \left| G^{v,i'}_{CO}(0,0)\right|^2}
     \:,
  \label{equ:cross_polar_level_multi}
\end{equation}
and
\begin{equation}
 DR(\theta)
  \equiv \frac{1}{2} \max_\phi
  \left|
  \frac{\sum_{i'} \left| G_{CO}^{v,i'}(\theta,\phi)\right|^2}
  {\sum_{i'} \left|G_{CO}^{v,i'}(0,0)\right|^2}
  -
  \frac{\sum_{i'}\left| G_{CO}^{h,i'}(\theta,\phi)\right|^2}
  { \sum_{i'}\left| G_{CO}^{h,i'}(0,0)\right|^2}
  \right|
  .
  \label{equ:differential_response_multi}
\end{equation}
The only difference from a single-mode system is 
the use of Eqs.~(\ref{equ:mm_power_s_matrix}),
(\ref{equ:s_matrix_vs_ampl_multimode_2x2matrix}) and
(\ref{equ:multi_mode_S_vs_G}),
as opposed to Eqs.~(\ref{equ:single_moded_response}),
(\ref{equ:b_vs_S_vs_a_in_single_mode}) and
(\ref{equ:single_mode_S_vs_G}),
in calculating the response powers
used in these definitions.

\subsection{Simplification for Infinitely Large,
  Electrically Thin Resistive
  Absorbers} \label{subsec:simpl_infty_large_absorber}
\begin{figure}[tbp]
 \begin{center}
  \includegraphics[width=0.48\textwidth]{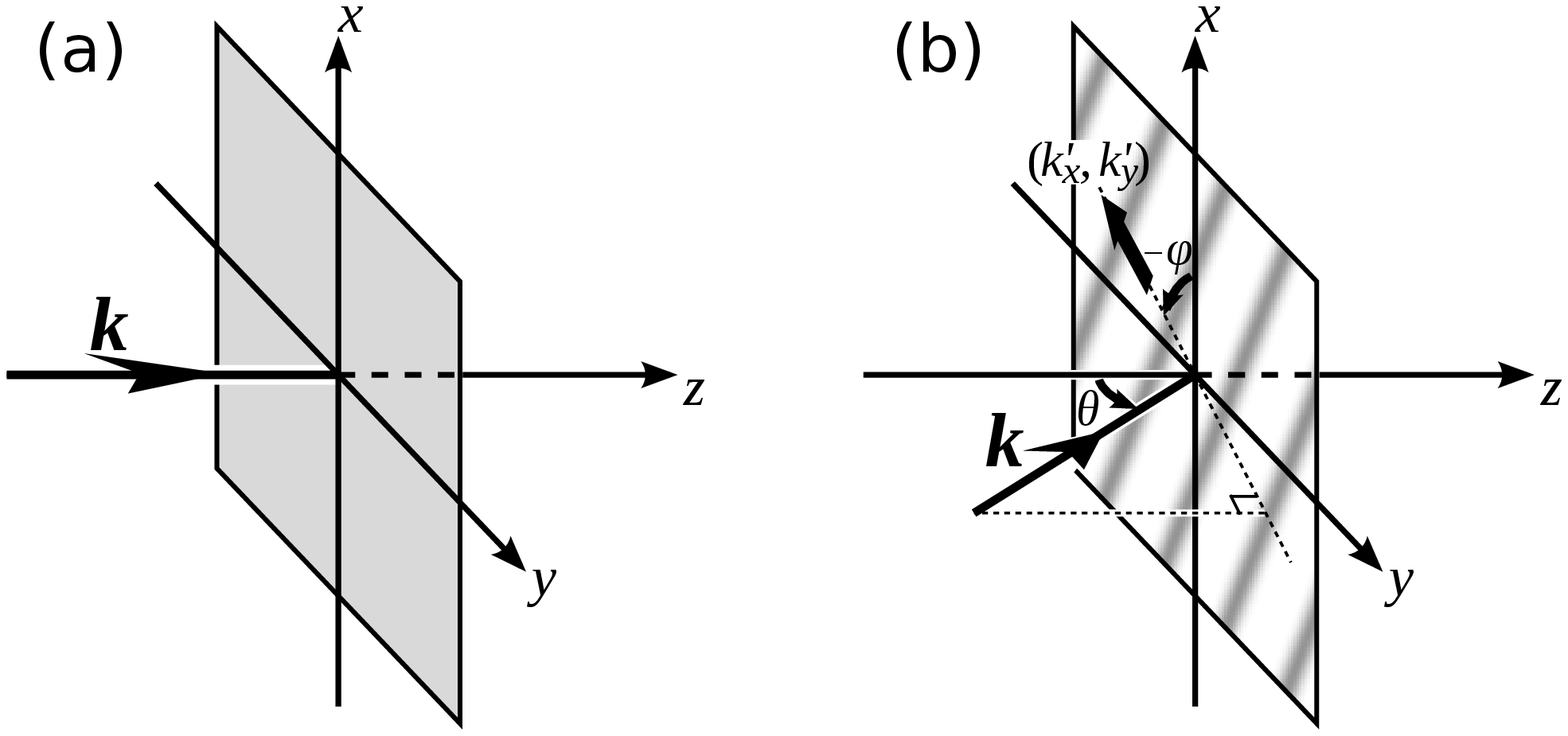}
  \caption{ \label{fig:surface_current_schematic} 
  Schematic figures showing the relation between the incident plane wave
  with a wave vector $\vect{k} \equiv (k_x, k_y, k_z)$
  and the induced surface current on an infinitely large absorber sheet.
  For an on-axis incident plane wave (a), the electromagnetic
  field on the absorber surface is spatially uniform 
  and thus the amplitude of the current is also uniform,
  while the direction of 
  the current is determined by the
  polarization of the incident wave.
  For an off-axis incident wave (b), on the other hand,
  the surface-current amplitude varies across the absorber
  sheet, which can be expressed in terms of two-dimensional Fourier modes
  with wave vectors
  $(k'_x, k'_y) \equiv (\tilde{k}\cos \phi', \tilde{k}\sin \phi')$
  where $\phi'$ is the polar-coordinate angle.
  Note that the amplitude propagation direction, $(k'_x, k'_y)$, is
  independent of
  the direction of the current, $(j_x, j_y)$;
  the former couples to the incident angle of the electromagnetic wave
  while the latter couples to its polarization.
  For an infinitely large absorber, the wavelength of the current variation
  must be as long or longer than the wavelength of the incident wave
  $(\tilde{k} \leq k)$, allowing us to express $\tilde{k}$
  as $\tilde{k} \equiv k \cos \theta'$.
  Further, the incident electromagnetic field on the $x$-$y$ plane has a
  wave vector of $(k_x,k_y)$ and thus only couples to
  the current mode with
  $(k'_x,k'_y) = (k_x,k_y)$.  In turn, this current mode
  only induces the electromagnetic field that has the wave vector
  $(k_x,k_y)$.  Thus, the coupling between the incident wave and the
  surface current
  is non-zero only for the Fourier mode of $(k'_x,k'_y)=(k_x,k_y)$,
  or $(\theta',\phi') = (\theta,\phi)$.
  }
 \end{center}
\end{figure}

Here, we relate the far-field functions in a multimoded system
$G_{CO[XP]}^{p,i'}(\theta,\phi)$
with the coupling coefficients $\widetilde{G}_{CO[XP]}^{p}(\theta,\phi)$
in Eq.~(\ref{equ:G__ampl_vs_current}) for an infinitely large thin
  resistive absorber,
and show the equivalence between the
$\widetilde{G}_{CO[XP]}^{p}(\theta,\phi)$
and the far-field functions for a single-mode system,
$G_{CO[XP]}^{p}(\theta,\phi)$.
The surface current on the absorber sheet
is denoted by its two-dimensional
Fourier amplitudes, $j_x(k'_x,k'_y)$ and $j_y(k'_x,k'_y)$,
where $k'_x$ and $k'_y$ are the wave numbers of the surface current,
and $j_x$ and $j_y$ correspond to 
current density per unit area of absorber in the
$x$ and $y$ directions, respectively.
For later convenience, we identify the Fourier modes 
using angular variables $(\theta',\phi')$
through
 $(k'_x,k'_y) \equiv (\tilde{k}\cos\phi',\tilde{k}\sin\phi')$
and
$\tilde{k} \equiv k \cos \theta'$ (see
Fig.~\ref{fig:surface_current_schematic});
we relabel the current amplitudes as
$j_p(k'_x,k'_y) \rightarrow j_p(\theta',\phi')$
accordingly.
We adopt the basis set of detector-coupled modes
that maps directly to the Fourier modes of the surface
current and use $(\theta',\phi')$ 
as the index $i'$ in
Eq.~(\ref{equ:s_matrix_vs_ampl_multimode_2x2matrix}):
\begin{equation}
  \vect{b}(\theta',\phi')
   \equiv
  \left(
  \begin{array}{c}
   b_v(\theta',\phi')\\
   b_h(\theta',\phi')
  \end{array}
  \right)
  \equiv
  \left(
  \begin{array}{c}
   \rho_x j_x(\theta',\phi')\\
   \rho_y j_y(\theta',\phi')
  \end{array}
  \right) \:,
  \label{equ:b_amplitude_for_sheet_vs_current}
\end{equation}
where $\rho_x$ and $\rho_y$ are the resistivity
of the absorber in the $x$ and $y$ directions, respectively.
We also relabel the $S$ matrix,
 $\vect{S}^{i'}(\theta,\phi) \rightarrow \vect{S}(\theta',\phi'; \theta,\phi)$.
Recall that we omit possible frequency dependence and thus the
discussion here is confined in a single wave number $k$.
The detected power $P_{\tilde{p}}$ for polarization $\tilde{p}$
is calculated through Eq.~(\ref{equ:mm_power_s_matrix}) by substituting
these definitions:
\begin{equation}
 P_{\tilde{p}} = \int \! \mathrm{d}\Omega'
  \rho_{\tilde{p}} \left|j_{\tilde{p}}(\theta',\phi')\right|^2
  = \frac{1}{\rho_{\tilde{p}}}
  \int \! \mathrm{d}\Omega'
  \left|b_p(\theta',\phi')\right|^2
  \:,
  \label{equ:thin_sheet_power_vs_b_ampl}
\end{equation}
with
$\mathrm{d}\Omega' \equiv \sin \theta' \, \mathrm{d}\phi' \,
\mathrm{d}\theta'$
and $(p,\tilde{p}) = (v,x)$ or $(h,y)$.
We also rewrite Eq.~(\ref{equ:s_matrix_vs_ampl_multimode_2x2matrix})
as
\begin{equation}
 \vect{b}(\theta',\phi') = 
  \int_\Omega \! \mathrm{d}\Omega \,
  \vect{S}(\theta',\phi';\theta,\phi) \vect{a}(\theta,\phi)
  \:.
  \label{equ:amplitudes_S_mapping_for_shin_absorber}
\end{equation}

We comment on our choice of the basis set of the
detector-coupled modes.
In order for Eq.~(\ref{equ:mm_power_s_matrix}) to be
valid,
the basis set
has to be the eigenmodes in dissipation process
in the device of interest.
A trivial choice here is the mode set
where each mode couples to
spatially localized surface current.
Our choice to use the Fourier modes is equivalent to the trivial choice
according to
Parseval's theorem (or the unitarity of Fourier transformation).

As shown in Fig.~\ref{fig:surface_current_schematic}
and discussed in Sec.~\ref{sec:elemag_calc}, an incoming mode with
an incident angle $(\theta,\phi)$ only induces surface currents with a
wave vector of
$(k'_x, k'_y) = (k\cos \theta\cos\phi,k \cos\theta\sin\phi)$.
Thus, the $S$ matrix can be written in a
simplified form as
\begin{equation}
 \vect{S}(\theta', \phi'; \theta, \phi)
  = 
  \delta(\cos \theta'- \cos \theta) \, \delta (\phi'-\phi)
   \, \vect{\widetilde{S}}(\theta, \phi) \:,
  \label{equ:s_matrix_and_delta}
\end{equation}
which is a product of 
a $2\times 2$ matrix $\vect{\widetilde{S}}$ that is
only dependent on $(\theta, \phi)$, and
the angular part of three-dimensional Dirac's delta function
in spherical coordinate system.
Equation~(\ref{equ:amplitudes_S_mapping_for_shin_absorber}) now reduces
to
\begin{equation}
 \vect{b}(\theta, \phi)
 =
 \vect{\widetilde{S}}(\theta, \phi) \, \vect{a}(\theta,\phi)
 \:.
 \label{equ:s_matrix_ampl_vs_current}
\end{equation}
Substituting Eq.~(\ref{equ:b_amplitude_for_sheet_vs_current})
into Eq.~(\ref{equ:s_matrix_ampl_vs_current}),
we obtain
\begin{equation}
 \left(
  \begin{array}{c}
   \rho_x j_x(\theta, \phi) \\
   \rho_y j_y(\theta, \phi)
  \end{array}
 \right)
 =
 \vect{\widetilde{S}}(\theta, \phi) \,
 \left(
  \begin{array}{c}
   a_v(\theta,\phi) \\
   a_h(\theta,\phi)
  \end{array}
 \right) \:.
 \label{equ:S__ampl_vs_current}
\end{equation}

Comparing Eq.~(\ref{equ:S__ampl_vs_current})
with Eq.~(\ref{equ:G__ampl_vs_current}),
one realizes the elements of the matrix $\vect{\widetilde{S}}$ are equal
to the coupling coefficients previously defined:
\begin{equation}
 \vect{\widetilde{S}}(\theta,\phi)
  =
  \left(
   \begin{array}{cc}
    \widetilde{G}_{CO}^{v}(\theta,\phi) & \widetilde{G}_{XP}^{v}(\theta,\phi) \\
    \widetilde{G}_{XP}^{h}(\theta,\phi) & \widetilde{G}_{CO}^{h}(\theta,\phi)
   \end{array}
  \right) \:.
  \label{equ:s_matrix_vs_far_field_function_thin_absorber}
\end{equation}
The far-field functions
$G_{CO [XP]}^{p, (\theta', \phi')}(\theta,\phi)$
 are related to the $S$ matrix via
Eq.~(\ref{equ:multi_mode_S_vs_G}), where the index $i'$ in the equation
is now labeled by $(\theta', \phi')$,
and thus related to the coupling coefficients through
Eqs.~(\ref{equ:s_matrix_and_delta})
and (\ref{equ:s_matrix_vs_far_field_function_thin_absorber}) as
\begin{equation}
 \begin{split}
  G_{CO[XP]}^{p, (\theta', \phi')} & (\theta,\phi)
  \\
   = 
  \delta & (\cos \theta'- \cos \theta) \, \delta (\phi'-\phi)
  \, \widetilde{G}_{CO[XP]}^{p}(\theta,\phi)
  \:,
 \end{split}
  \label{equ:far_field_function_delta}
\end{equation}
As noted earlier, the $\widetilde{G}_{CO [XP]}^{p}(\theta,\phi)$ 
are not
 the far-field functions
 since they do not define a radiation field pattern by
 Eq.~(\ref{equ:radiation_field_by_Gco_Gxp}).
The far-field functions are defined by
the left hand side of Eq.~(\ref{equ:far_field_function_delta}),
 where each mode labeled by $(\theta', \phi')$
 has a radiation pattern
 corresponding to a single plane wave,
 and $\widetilde{G}_{CO [XP]}^{p}(\theta,\phi)$ 
 represents the relative excitation strengths
 among the modes.

Nevertheless, 
$\widetilde{G}_{CO [XP]}^{p}(\theta,\phi)$ 
hold a formal equivalence to the far-field functions
in a single-mode system, $G_{CO [XP]}^{p}(\theta,\phi)$.
For example,
substituting Eqs.~(\ref{equ:s_matrix_ampl_vs_current})
and (\ref{equ:s_matrix_vs_far_field_function_thin_absorber})
into 
Eq.~(\ref{equ:thin_sheet_power_vs_b_ampl})
leads to 
 an expression for the detected power
equivalent to that for a single-mode sensor:
\begin{equation}
 P_{\tilde{p}} =
 \frac{1}{\rho_{\tilde{p}}} 
  \int_\Omega \! \! \mathrm{d}\Omega 
  \left|
   \widetilde{G}_{CO}^{p}(\theta,\phi) a_p(\theta,\phi)
   + \widetilde{G}_{XP}^{p}(\theta,\phi) a_{\bar{p}}(\theta,\phi)
  \right|^2 ,
  \label{equ:power_of_thin_resistive_absoeber_vs_amplitudes}
\end{equation}
with $(\tilde{p},p,\bar{p}) = (x,v,h)$ or $(y,h,v)$.
Substituting Eq.~(\ref{equ:far_field_function_delta})
into Eqs.~(\ref{equ:cross_polar_level_multi}) and
(\ref{equ:differential_response_multi})
yields forms of $XP(\theta)$ and $DR(\theta)$
that are equivalent to
Eqs.~(\ref{equ:cross_polar_level}) and
(\ref{equ:differential_response})\footnote{Note that
the label for the detected modes is continuous variable,
$(\theta',\phi')$, in Eq.~(\ref{equ:far_field_function_delta}),
while it is discrete, $i'$, in Eqs.~(\ref{equ:cross_polar_level_multi}) and
(\ref{equ:differential_response_multi}).
For a consistent treatment, one should either replace the Dirac's
$\delta$ function in Eq.~(\ref{equ:far_field_function_delta})
with Kronecker's $\delta$ in the substitution,
or use Eq.~(\ref{equ:power_of_thin_resistive_absoeber_vs_amplitudes})
and the literal definitions of $DR(\theta)$ and $XP(\theta)$ described at
the end of Sec.~\ref{subsec:far_field_in_single_mode}.  Both treatments
lead to the same result equivalent to Eqs.~(\ref{equ:cross_polar_level}) and
(\ref{equ:differential_response}).
}.
This formal equivalence
to a single-mode system
allows us to regard $\widetilde{G}_{CO[XP]}^{p}(\theta,\phi)$
in Eq.~(\ref{equ:G__ampl_vs_current}), or 
in Eq.~(\ref{equ:far_field_function_delta}),
equivalently to the far-field functions in a single-mode system.
Note that Eq.~(\ref{equ:power_of_thin_resistive_absoeber_vs_amplitudes})
can be generalized to represent the response to an arbritary incoming
distribution, which can be expressed as a spectrum of plane waves.

We note that our choice of
identifying the Fourier modes by $(\theta',\phi')$
has simplified the derivation above,
allowing us to evade possible Jacobians
that would arise with an other choice, e.g., 
$(k'_x,k'_y)$.
Our choice is convenient since,
after relating $(\theta',\phi')$ to $(\theta,\phi)$ through Dirac's
delta function, the power measured by the detector can be calculated
through an integral over the solid angle,
$P = \int \! \mathrm{d}\Omega \, \rho |j|^2$, as is standard.
This has allowed us to directly equate the 
matrix $\vect{\widetilde{S}}(\theta,\phi)$
with the $\widetilde{G}_{CO [XP]}^{p}(\theta,\phi)$
defined in Eq.~(\ref{equ:G__ampl_vs_current}),
which are independently introduced;
the former is introduced in describing the $S$ matrix
while the latter is the coupling coefficients
in the process where
a single plane wave interacts with the sheet absorber
and induces the surface current.
We also note that the derivation above has clarified the modal content of
the current $j_p$ in Eq.~(\ref{equ:G__ampl_vs_current})
and its dependence on the incident angle $(\theta,\phi)$;
induced current is a pure Fourier mode with a wave vector
$(k'_x,k'_y) = (k\cos\theta\cos\phi, k\cos\theta\sin\phi)$.

\section{Response Pattern of an Infinitely Large, Electrically Thin Resistive
 Absorber} \label{sec:elemag_calc}
In this section, we calculate the response patterns for an infinitely
large, electrically thin resistive absorber.
The absorber sheet may comprise either an electrically thin resistive membrane
or a pair of orthogonal grids,
where the latter could be constituted as a dual-polarization polarimeter by thermally
coupling each of the grids to a power sensor.
In the following, we treat the resistive grid-pair absorber equivalently to a
membrane absorber
regarding its electromagnetic coupling to free space,
except that a grid-pair is more general in that it may have
different
resistivities in the two
orthogonal directions.  The equivalence is based on the assumptions that
the grid
 is thin and finely pitched, and  the
 distance between the stacked grids is small.
Discussions on concrete conditions and 
  the validity of the
  treatment can be found in
 Appendix~\ref{app:detector_geometry_assumption}.
Hereafter, we assume the more general case.
We define the coordinates such that the absorber sheet is parallel to the
$x$-$y$ plane,
where each absorber grid is aligned in the $x$ or $y$ direction,
and the wave vector of an incoming plane wave
is
$\vect{k}\equiv (k_x,k_y,k_z)^T$ with $\left|\vect{k}\right|\equiv k$.
Note that, although the calculation in this section assumes a plane-wave
incident mode, the discussion in the previous section allows us to apply
the calculated result to an arbitrary incident mode through
Eq.~(\ref{equ:power_of_thin_resistive_absoeber_vs_amplitudes}).
This is of importance since, in most of uses, the sensor is at the focus
of optics, where the mode incident on the sensor is not a plane wave.
Appendix~\ref{app:koffman_relation} also discusses a case where the incident mode is not a plane
wave.

In the following, we first discuss the case of a freestanding thin
absorber.  Although this is
a configuration that minimizes the polarization systematics, the maximum
absorption efficiency is limited to 50\% as in an analogous case of 
the transmission line (see also 
Fig.~\ref{fig:no_bs_transmission_line} in Appendix~\ref{sec:appendix_calc_without_backshort}).
We then discuss the case that employs a reflective backshort, which is
a technique 
frequently used to improve the absorption efficiency
(see also Fig.~\ref{fig:with_bs_transmission_line} in
Appendix~\ref{sec:appendix_calc_with_backshort}).  As we will show, it
significantly improves the efficiency while introducing small systematics.

\subsection{Freestanding Electrically Thin Absorber \label{sec:no_backshort}}
\begin{figure}[tbp]
 \begin{center}
  \includegraphics[width=0.3\textwidth]{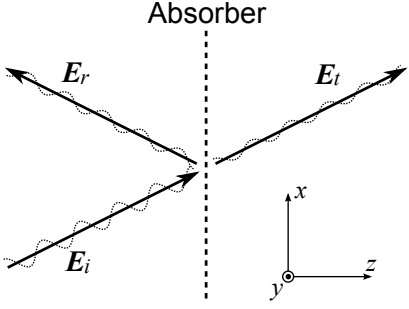}
  \caption{\label{fig:no_backshort} 
  The geometric arrangement of the incident, reflected, and transmitted
  fields and the thin resistive absober sheet placed at $z=0$.
  }
 \end{center}
\end{figure}

We consider an arrangement shown in Fig.~\ref{fig:no_backshort}, where an
electrically thin resistive absorber is at $z=0$, lying on the $x$-$y$ plane.
The electric field of an incoming plane wave
with arbitrary polarization
can be expressed as a linear combination of 
vertically and holizontally polarized components with amplitudes
$a_v \equiv a_v(\theta, \phi)$ and
 $a_h \equiv a_h(\theta, \phi)$:
\begin{equation}
 \vect{E}_i(\vect{r};\vect{k})
   = e^{-i \vect{k}\cdot \vect{r}}
  \left(
   a_v \ev + a_h \eh
  \right) \:.
\end{equation}
The transmitted wave ($\vect{E}_t$) and reflected wave ($\vect{E}_r$) have
wave vectors of $\vect{k}$ and
$\vect{\bar{k}} \equiv (k_x,k_y,\pm k_z)^T$.
At $z=0$, all three plane wave components have the same phase of
$e^{-i(k_x x +k_y y)}$ and only the surface current mode
with a wave vector of $(k'_x, k'_y) = (k_x, k_y)$ is induced.
Thus, the current density of the induced current is written as
\begin{equation}
 \vect{J}(x,y,z) = \delta(z) e^{-i(k_x x +k_y y)}
  \left( j_x \ex + j_y \ey \right) \:,
\end{equation}
where $j_x$ and $j_y$ are the Fourier amplitudes of the surface
current per unit absorber area.  In turn, this current only induces
radiating
electromagnetic waves
with the phase of $e^{-i(k_x x +k_y y)}$  at $z=0$
(Appendix~\ref{sec:appendix_calc_without_backshort}), 
and thus the $\delta$ functions in Eq.~(\ref{equ:s_matrix_and_delta})
are confirmed.

\begin{figure}[tbp]
 \includegraphics[width=0.45\textwidth]{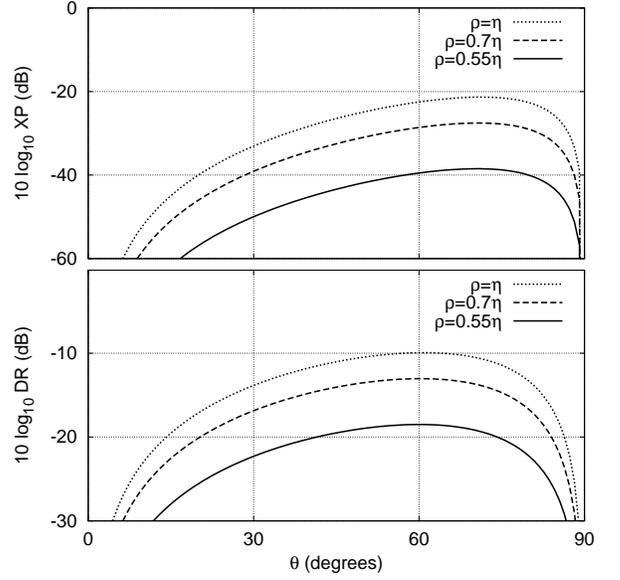}
 \caption{\label{fig:xp_dr_no_bs} 
 Cross-polar level (top) and differential response (bottom) of 
 an electrically thin, infinitely large resistive absorber
 with $\rho = \eta$, $0.7\eta$, and $0.55\eta$.
 Both are zero when $\rho = \eta / 2$.
 One can see the approximate order of magnitude relation of
 $DR \sim  \sqrt{XP}$ holds here.
 }
\end{figure}

Solving boundary conditions (see
Appendix~\ref{sec:appendix_calc_without_backshort}),
we 
derive the coupling coefficients in Eq.~(\ref{equ:G__ampl_vs_current})
relating $(j_x, j_y)$ and $(a_v, a_h)$:
\begin{equation}
 \left(
  \begin{array}{cc}
   \widetilde{G}^v_{CO}/\rho_x & \widetilde{G}^v_{XP}/\rho_x \\
   \widetilde{G}^h_{XP}/\rho_y & \widetilde{G}^h_{CO}/\rho_y
  \end{array}
 \right)^{-1}  =
  \frac{1}{4 \cos \theta}
 \left(
  \begin{array}{cc}
   \alpha_+ (\rho_x) & \beta (\rho_y) \\
   \beta (\rho_x) & \alpha_- (\rho_y) \\
  \end{array}
 \right)
 \label{equ:sigma_matrix_no_bs}
\end{equation}
with
\begin{equation}
 \begin{split}
 \alpha_\pm (\rho)
  & \equiv
  (2 \rho + \eta) (1 + \cos \theta) \pm 
  \cos 2 \phi (2 \rho - \eta) (1 - \cos \theta) \:,
  \\
  \beta (\rho)
  & \equiv
  \sin 2 \phi (2 \rho - \eta) (1 - \cos \theta)
  \:,
 \end{split}
 \label{equ:alpha_beta_no_bs}
\end{equation}
where $\eta$ is the impedance of free space.
When the resistivities are the same for the $x$ and $y$ directions,
 $\rho_x = \rho_y \equiv \rho$, the cross-polar level and the
 differential response defined by Eqs.~(\ref{equ:cross_polar_level}) and
 (\ref{equ:differential_response}) are
\begin{equation}
 \begin{split}
 XP&(\theta)
  \\
  &=
  \frac{1}{4}\cos^2\theta (1-\cos \theta)^2
  \frac{(2\rho-\eta)^2 (2\rho+\eta)^2}
  { (2\rho + \eta \cos \theta)^2 (2\rho \cos \theta +\eta)^2}
 \end{split}
 \label{equ:xp_no_bs}
\end{equation}
and
\begin{equation}
  DR(\theta)
   =
  \frac{1}{8}\sin^2 2 \theta
  \frac{|2\rho-\eta| (2\rho+\eta)^3}
  { (2\rho + \eta \cos \theta)^2 (2\rho \cos \theta +\eta)^2}
  \:.
 \label{equ:dr_no_bs}
\end{equation}
Figure~\ref{fig:xp_dr_no_bs} shows $XP(\theta)$ and $DR(\theta)$
for various $\rho$.

The most important implication is the case with
$\rho_x = \rho_y = \eta/2$.
As can be seen from Eqs.~(\ref{equ:sigma_matrix_no_bs}),
(\ref{equ:alpha_beta_no_bs}), (\ref{equ:S__ampl_vs_current}), and
(\ref{equ:s_matrix_vs_far_field_function_thin_absorber}),
this is
the necessary and sufficient condition for
the matrix relating $(j_x, j_y)$ and $(a_v, a_h)$ to be diagonal,
\begin{equation}
 \left(
  \begin{array}{c}
   j_x \\
   j_y \\
  \end{array}
 \right)
  = \frac{2\eta^{-1} \cos \theta}{1+\cos \theta}
 \left(
  \begin{array}{c}
   a_v \\
   a_h \\
  \end{array}
 \right),
\end{equation}
and for Eq.~(\ref{equ:zero_systematic_condition}) to be satisfied
 for $\widetilde{G}_{CO [XP]}^{p}(\theta,\phi)$.
Thus, this surface resistivity
 achieves zero cross polarization and zero differential response,
 $XP(\theta) = DR(\theta) = 0$.  We discuss the underlying symmetry that
 causes the systematics to vanish later in this section.

\begin{figure}[tbp]
 \begin{center}
  \includegraphics[width=0.45\textwidth]{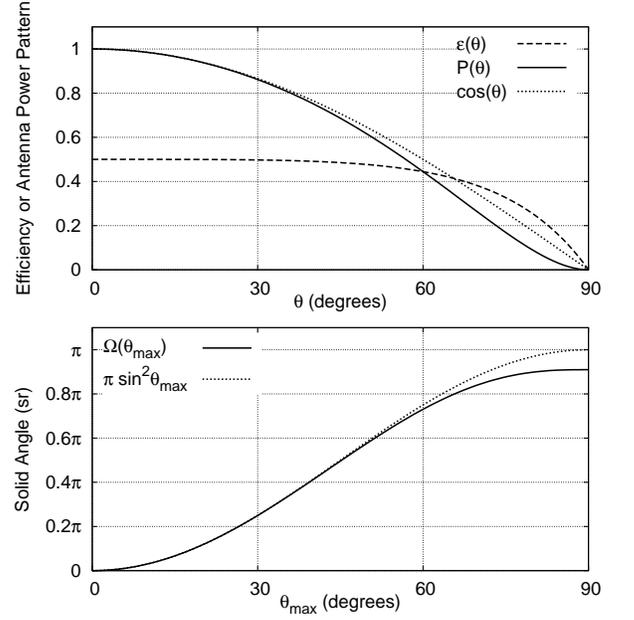}
  \caption{\label{fig:beam_shape} 
  Top: the absorption efficiency $\epsilon(\theta)$ 
  and the antenna power pattern $P(\theta)$
  of a large electrically thin resistive absorber with surface
  resistivity $\rho = \eta/2$
  [see Eqs.~(\ref{equ:antenna_power_pattern_def}) and (\ref{equ:efficiency_def})].
  Also shown for comparison is the 
  antenna power pattern for constant absorption efficiency, $\cos \theta$,
  na\"{i}vely expected from geometry.
  Bottom: the solid angle $\Omega$ as a function of maximum acceptance
  $\theta_{\mathrm{max}}$ for the thin resistive absorber 
  with $\rho = \eta/2$.
  Compared is the na\"{i}ve geometric expectation of
  $\pi \sin^2 \theta_{\mathrm{max}}$.
  }
 \end{center}
\end{figure}

In this special case of $\rho_x = \rho_y = \eta/2$, the angular
dependence of the power
absorption per unit area
of the resistive surface, $P_a$, is
\begin{equation}
 P_a(\theta, \phi) = P_a(\theta) = \rho \, j^2 = 
  \frac{2\eta^{-1} \cos^2 \theta}{(1+\cos \theta)^2} \, a^2 \:,
  \label{equ:power_absorption_no_bs}
\end{equation}
where $(\rho, j, a)$ represents either $(\rho_x, j_x, a_v)$ or 
$(\rho_y, j_y, a_h)$.  The power flow of the incident
plane wave per unit area per polarization is $P_i = \eta^{-1} a^2$.
Thus, the angular
dependence of the absorption efficiency is
\begin{equation}
 \epsilon(\theta) \equiv \frac{P_a(\theta)}{\cos \theta \, P_i}
  = \frac{2 \cos \theta}{(1+\cos \theta)^2} ,
  \label{equ:efficiency_def}
\end{equation}
where the extra factor of $\cos \theta$ arises in converting the power
absorption $P_a(\theta)$ to be per unit area of incident wave.
The efficiency is maximum at 50\% at $\theta=0$, as
expected from an analogous transmission line configuration
(see Fig.~\ref{fig:no_bs_transmission_line} in
Appendix~\ref{sec:appendix_calc_without_backshort}),
and slowly drops as $\theta$ increases since the effective surface
resistivity for an off-axis incident wave deviates from $\eta/2$.
The antenna reception power pattern $P(\theta)$ is obtained by
normalizing
$P_a(\theta)$ to its maximum:
\begin{equation}
 P(\theta) \equiv \frac{P_a(\theta)}{\max_\theta P_a(\theta)} = \frac{4 \cos^2 \theta}{(1+\cos
  \theta)^2} \:,
  \label{equ:antenna_power_pattern_def}
\end{equation}
where $\max_\theta P_a(\theta) = P_a(0)$ in this case.
Figure~\ref{fig:beam_shape} (top) shows the efficiency
$\epsilon(\theta)$ 
and the antenna power pattern $P(\theta)$.
Since $\epsilon(\theta)$ is nearly constant,
  $P(\theta)$ is 
close to $\cos \theta$, the na\"{i}ve expectation due to the
geometric factor.
  By integrating
$P(\theta)$, we obtain the solid angle as a function of the maximum
acceptance
angle $\theta_{\mathrm{max}}$:
\begin{equation}
 \begin{split}
  \Omega(\theta_{\mathrm{max}})\equiv & \, 2\pi \int^{\theta_{\mathrm{max}}}_0
  d(\cos \theta) P(\theta)
  \\
   = & \, 4\pi \Bigl[
	  \frac{(3+2\cos
  \theta_{\mathrm{max}})(1-\cos\theta_{\mathrm{max}})}{1+\cos
  \theta_{\mathrm{max}}}
  \\
  & \quad  \quad \quad  \quad  \quad  \quad  \quad 
  + 4 \log \frac{1+\cos \theta_{\mathrm{max}}}{2}
	 \Bigr] .
 \end{split}
 \label{equ:solid_angle_wo_bs}
\end{equation}
Figure~\ref{fig:beam_shape} (bottom) shows $\Omega(\theta_{\mathrm{max}})$ in comparison to
the na\"{i}ve geometric expectation of $\pi \sin^2 \theta_{\mathrm{max}}$.  For
a maximum
acceptance at the $\theta_{\mathrm{max}}=\pi/2$, 
$\Omega \simeq 0.91 \pi$ while the
na\"{i}ve geometric expectation yields $\pi$.

\begin{figure}[tbp]
 \includegraphics[width=0.42\textwidth]{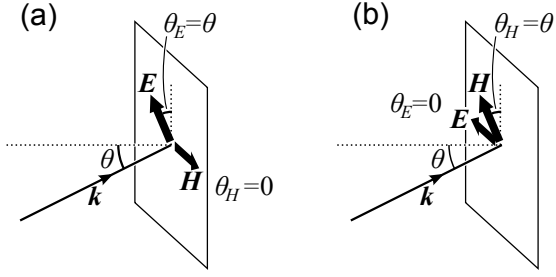}
 \caption{ \label{fig:e_plane_vs_h_plane_incident} 
 Schematic figure of (a)
 $E$-plane and (b) $H$-plane
 incident waves with wave vector $\vect{k}$.
 The former (latter) has its electric (magnetic) field parallel to the
 plane of incidence.
 Shown are the relation among
 the incident angle ($\theta$),
 the angle between $\vect{E}$ field and the absorber plane ($\theta_E$),
 and the angle between $\vect{H}$ field and the absorber plane
 ($\theta_H$).
 }
\end{figure}

\begin{figure}[tbp]
 \includegraphics[width=0.47\textwidth]{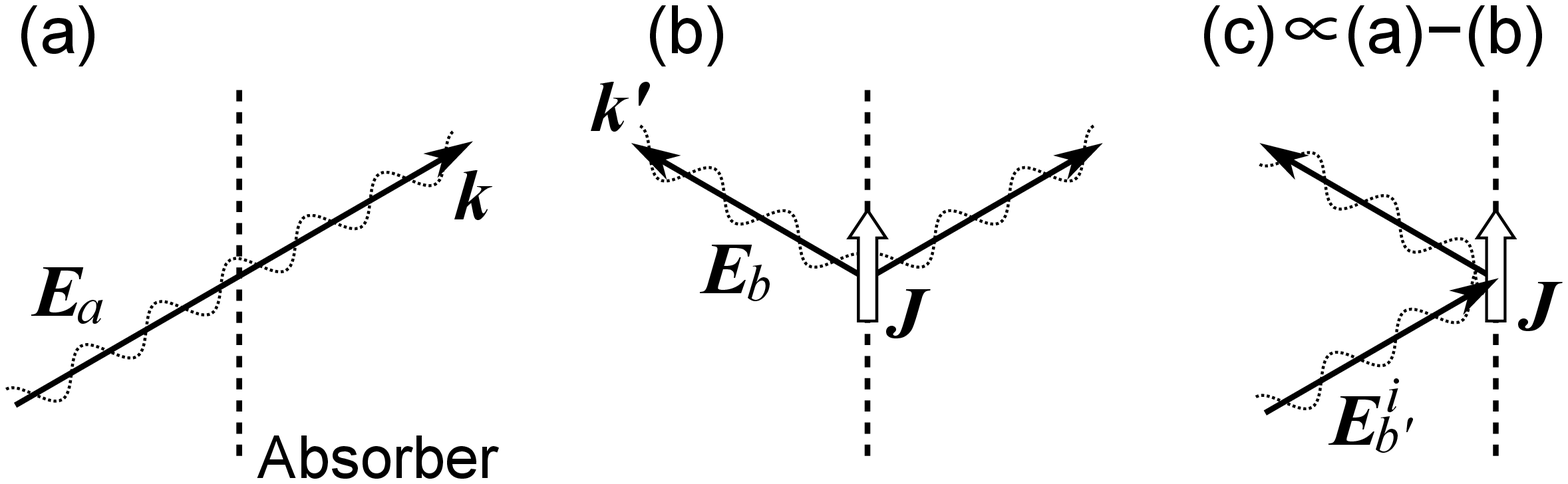}
 \caption{\label{fig:boundary_condition} 
 Two ways to form a solution for a resistive absorber sheet
 interacting with an incident plane wave.
 One is to form a linear combination of (a) and (b).
 Here, (a) is the
 plane wave propagating without any interaction,
 or the limit of highly resistive absorber ($\rho \gg \eta$),
 and (b) is the current on the
 sheet
 and the current-induced radiating fields.
 The boundary condition, or Ohm's law, constrains 
 the coefficients of the linear combination by
 $\left(\vect{E}_a + \vect{E}_b \right)\bigr|_s = \rho \vect{J}$,
 where $\vect{E}\bigr|_s$ denotes the electric field projected on the
 plane of the absorber sheet.
 Another way to form the solution
 is to use a combination of (a) and (c),
 instead of the combination of (a) and (b).
 Here, (c) corresponds to a
 solution for the case
 when the sheet is a perfect conductor,
 as described in the text.
 A linear combination of (a) and (c) can form the solution 
  since (c) is a linear combination of (a) and (b).
 }
\end{figure}

It may appear non-trivial that the response of an 
electrically thin resistive grid or membrane
completely eliminates polarization systematics when $\rho = \eta/2$.
One can intuitively understand why the response pattern attains
symmetry between the $E$- and $H$-planes (leading to zero
systematics) as follows.
As shown in Fig.~\ref{fig:e_plane_vs_h_plane_incident},
an $E$-plane ($H$-plane) incident wave is a linearly-polarized plane
wave with its electric field parallel (perpendicular) to the plane of incidence.
When $\rho_x = \rho_y = \rho$, the $E$- and $H$-plane coupling
coefficients,
$\widetilde{G}_{CO}^v(\theta,\phi=0)$ and
$\widetilde{G}_{CO}^h(\theta,\phi=0)$,
relate the amplitudes of
 the incident field and
 the induced current
as
\begin{equation}
 \begin{split}
 a_v & = \frac{1}{\cos \theta}\left(\rho  + \frac{\eta}{2} \cos \theta\right) j_x \:,
  \\
  a_h & = \frac{1}{\cos \theta}\left(\rho \cos \theta + \frac{\eta}{2} \right) j_y \:,
 \end{split}
 \label{equ:h_vs_e_plane_current_vs_incident}
\end{equation}
where $a_v$ ($a_h$) is the electric field amplitude for $E$-plane
($H$-plane) incident since $\ev = \eq$ ($\eh=\ep$) at $\phi=0$.
One can see each of Eq.~(\ref{equ:h_vs_e_plane_current_vs_incident})
consists of
two terms.
The first term dominates in the limit
$\rho \gg \eta$ (thin film resistor approximates an open) and the second
dominates in the limit
 $\rho \ll \eta$ (termination approximates a short).  At the limit of
$\rho \gg \eta$ (Fig.~\ref{fig:boundary_condition}a), the
electromotive force and current-induced field can be ignored.
In this limit, the current
is simply proportional to the component of the electric field parallel
to the resistive surface, and written as
\begin{equation}
 \rho \left|\vect{J}\right| =  \cos \theta_E  \left| \vect{E}_i \right|
  \label{equ:voltage_drop_due_to_resistivity}
\end{equation}
where $\theta_E$ is the angle between the electric field and the
absorber surface. 
Figure~\ref{fig:e_plane_vs_h_plane_incident} defines $\theta_E$ such that
$\theta_E = \theta$ ($\theta_E = 0$) for $E$-plane ($H$-plane) incident,
and thus Eq.~(\ref{equ:voltage_drop_due_to_resistivity}) is in agreement
with the first terms of Eq.~(\ref{equ:h_vs_e_plane_current_vs_incident}).
At the limit of $\rho \ll \eta$ (Fig.~\ref{fig:boundary_condition}c), on the other hand, the absorber
approximates a perfect electric conductor, and the field of the incident wave and the
current are
related as $\vect{J} = 2 \vect{n} \times \vect{H}_i$,
where $\vect{n}=-\ez$ is the unit vector normal to the absorber sheet.
Thus,
\begin{equation}
 \frac{\eta}{2} \left|\vect{J}\right| = \cos \theta_H \left| \vect{E}_i \right|
  \label{equ:electromotive_force}
\end{equation}
where $\eta  \left| \vect{H}_i \right|=  \left|\vect{E}_i \right|$
and $\theta_H$ is the angle between the magnetic field and the
absorber sheet surface.
For $E$-plane ($H$-plane) incident,
$\theta_H = 0$ ($\theta_H = \theta$) as shown in
Fig.~\ref{fig:e_plane_vs_h_plane_incident}.
When the value of $\rho$ is 
at neither of the limits,
the electric field simply equals to the sum of the voltage drop due
to resistivity, Eq.~(\ref{equ:voltage_drop_due_to_resistivity}), and the
electromotive force, Eq.~(\ref{equ:electromotive_force}); the electric field
of the incident wave is
related to the current as
\begin{equation}
 \left| \vect{E}_i \right|
  =
  \left( 
  \rho \frac{1}{\cos \theta_E} 
  + 
  \frac{\eta}{2} \frac{1}{\cos \theta_H} 
  \right)
  \left|\vect{J}\right| \:.
  \label{equ:j_vs_e_h_surface}
\end{equation}
This follows from the boundary condition in forming the solution
as a linear combination of the limited cases shown in Fig.~\ref{fig:boundary_condition}a and
\ref{fig:boundary_condition}c.
The total incident field is the sum of those for Fig.~\ref{fig:boundary_condition}a and \ref{fig:boundary_condition}c:
$\vect{E}_i = \vect{E}_a + \vect{E}_{b'}^i$.
For this linear combination, the electric field projected on the
absorber sheet is
simply that of $\vect{E}_a$ since it vanishes for Fig.~\ref{fig:boundary_condition}c.
Thus, Ohm's law corresponds to
$\vect{E}_a \bigr|_s = \rho \vect{J}$, where $\bigr|_s$ denotes a
projection on the plane of the absorber sheet.
On the other hand, the field in Fig.~\ref{fig:boundary_condition}c
satisfies $2\vect{n}\times\vect{H}_{b'}^{i}=\vect{J}$,
where $\vect{H}_{b'}^{i}$ is the magnetic field associated to
$\vect{E}_{b'}^i$.
Substituting these two conditions into 
$\vect{E}_i = \vect{E}_a + \vect{E}_{b'}^i$
leads to Eq.~(\ref{equ:j_vs_e_h_surface}).

Equation~(\ref{equ:j_vs_e_h_surface}) is equivalent to
Eq.~(\ref{equ:h_vs_e_plane_current_vs_incident}).
Now it is obvious why the current response to the incident wave
becomes symmetric between $E$- and $H$-planes when $\rho = \eta/2$,
since this condition makes Eq.~(\ref{equ:j_vs_e_h_surface}) symmetric to
the exchange of $E$ and $H$.
This also clarifies the observation and can be used to understand why a
wire-grid made from perfect conductor is not a perfect polarizer~\cite{2009stt..conf..223L}.

It is worth noting that the same angular response
as that obtained in the limit of $\rho \gg \eta$ would also show up if we
replaced the
resistive sheet by a perfect magnetic conductor.  In this sense,
zero systematics are attained when there are both electric and magnetic
conductor sheets and the contributions from the two
are equal.  This is exactly the
same symmetry as the one discussed by Koffman~\cite{1966ITAP...14...37K}
in the context of cross polarization of a feedhorn, where
the feedhorn coupled to a paraboloid reflector attains zero cross
polarization when the radiation pattern
can be described as a balanced sum of electric-dipole and
 magnetic-dipole radiation fields (Huygens source); see
 Appendix~\ref{app:koffman_relation} for details.  The symmetry between the
 electric and
 magnetic dipoles
discussed by Koffman
corresponds to that of electric and magnetic conductor sheets
discussed here.
We also note that our 
 work can be seen as an extension of
the current sheet model for phased-array
antennas~\cite{1965ITAP...13..506W,hansen2009phased}.
The model assumes
current sheet of surface impedance $\eta$ backed with a magnetic reflector
to simulate a phased array,
and derives the symmetry of antenna properties,
such as the reflection coefficent and scanning impedance,
between $E$- and $H$-planes.
It parallels to our discussion above
and its result is consistent with our derivation.
For example, it suggests that the absolute value of the reflection
coefficient is $\tan^2 \frac{\theta}{2}$ in both $E$- and
$H$-planes, while
Eq.~(\ref{equ:efficiency_def}) can be rewritten as
$\epsilon(\theta) = \left(1 - \tan^2 \frac{\theta}{2}\right)/2$.
The factor two differences
both in the absorption efficiency and in the resistivity
to achieve the symmetry
follow the expectation
due to the absence of the magnetic reflector in our setup; 
see our discussion at the beginning of this section pointing out the
analogy to the case of transmission 
line (cf. Fig.~\ref{fig:no_bs_transmission_line} in Appendix~\ref{sec:appendix_calc_without_backshort}) and the maximum
absorption efficiency of 50\% without a backshort.

\subsection{Electrically Thin Absorber with Reflective Backshort Termination}
\begin{figure}[tbp]
 \begin{center}
  \includegraphics[width=0.47\textwidth]{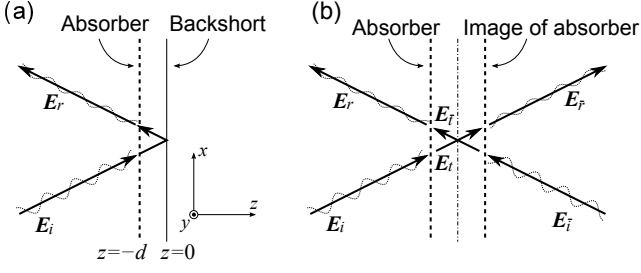}
  \caption{\label{fig:geometry_with_backshort} 
  (a) The geometric arrangement of the incident and reflected
  fields, the absober sheet placed at $z=-d$, and a
  reflective backshort placed at $z=0$.
  (b) The virtual arrangement exploiting the image theory,
  equivalent to (a) in the region of $z<0$.
  }
 \end{center}
\end{figure}

Use of a reflective backshort behind the electrically thin absorber sheet improves
the absorption efficiency and is desirable in practical applications.
We consider an arrangement shown in
Fig.~\ref{fig:geometry_with_backshort}a, where the absorber is
at $z=-d$ and a reflective backshort is at $z=0$, 
with both parallel to the $x$-$y$ plane.  In 
calculating the response pattern, we exploit the image theory and use
the arrangement of
Fig.~\ref{fig:geometry_with_backshort}b.
Solving boundary conditions (see
Appendix~\ref{sec:appendix_calc_with_backshort}),
the coupling coefficients 
are derived as
\begin{equation}
 \begin{split}
  & \left(
  \begin{array}{cc}
   \widetilde{G}^v_{CO}/\rho_x & \widetilde{G}^v_{XP}/\rho_x \\
   \widetilde{G}^h_{XP}/\rho_y & \widetilde{G}^h_{CO}/\rho_y
  \end{array}
  \right)^{-1}
  \\
  & \quad \quad \quad \quad \quad \quad = 
  \frac{e^{-i k_z d}}{4 \sin k_z d \cos \theta}
 \left(
  \begin{array}{cc}
   \alpha^{\mathrm{bs}}_+ (\rho_x) & \beta^{\mathrm{bs}} (\rho_y) \\
   \beta^{\mathrm{bs}} (\rho_x) & \alpha^{\mathrm{bs}}_- (\rho_y) \\
  \end{array}
 \right),
 \end{split}
\end{equation}
with
\begin{equation}
 \begin{split}
 \alpha^{\mathrm{bs}}_\pm (\rho)
  \equiv & 
  \left[ (\rho + \eta) \sin k_z d -  i \rho \cos k_z d \right]
  \left(1 + \cos \theta \right)
  \\
  & \quad \quad \pm \cos 2\phi \: \delta^{\mathrm{bs}}(\rho, d, \theta) \:,
  \\
  \beta^{\mathrm{bs}} (\rho)
  \equiv & \sin 2\phi \: \delta^{\mathrm{bs}}(\rho, d, \theta)
   \:,
 \end{split}
 \label{equ:alpha_beta_with_backshort}
\end{equation}
and
\begin{equation}
 \delta^{\mathrm{bs}}(\rho, d, \theta)
  \equiv
  \left[
  (\rho - \eta) \sin k_z d - i \rho \cos k_z d
  \right]
  \left( 1-\cos \theta \right) \:.
  \label{equ:delta_with_backshort}
\end{equation}
This equation is already instructive even without calculating the
cross-polar level and the differential response explicitly.  In order to
eliminate these systematics, the matrix elements should 
satisfy $\alpha_+^\mathrm{bs} = \alpha_-^\mathrm{bs}$
and $\beta^\mathrm{bs} = 0$.  This is only possible if
$\delta^{\mathrm{bs}} (\rho, d, \theta) = 0$.  Even though
the first term in $\delta^{\mathrm{bs}} (\rho, d, \theta)$
vanishes for $\rho=\eta$,  
$\delta^{\mathrm{bs}} (\rho, d, \theta)$ cannot be zero for an arbitrary
$\theta$ since $k_z = k \cos \theta$.
Thus, there always are non-zero systematics, unlike the case without a
backshort.
However, by optimizing the distance
to the backshort $d$, a low level of beam systematics can still be achieved.

\begin{figure}[tbp]
 \includegraphics[width=0.45\textwidth]{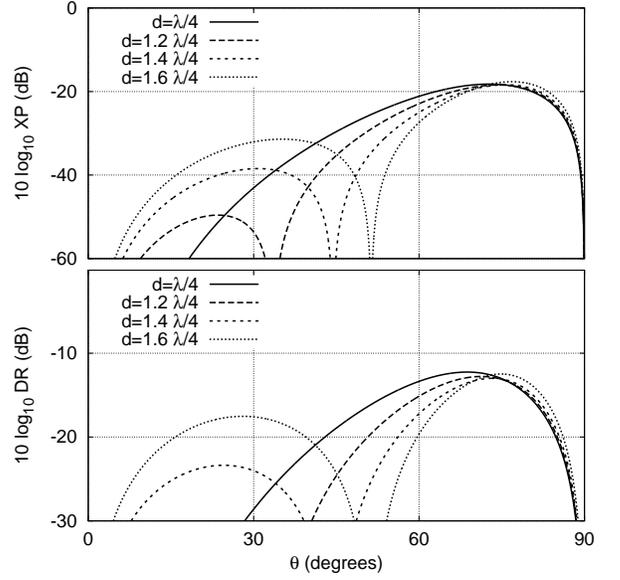}
 \caption{\label{fig:xp_dr_with_bs} 
 Cross-polar level (top) and differential response (bottom) of electrically thin
 resistive sheet absorption
 pattern with reflective backshort,
 for $d=\lambda/4$, $d=1.2 \lambda/4$, $d=1.4 \lambda/4$,
 and $d=1.6 \lambda/4$, where $\lambda$ and $d$ are the wavelength and
 the distance between the sheet and the reflective backshort.
 }
\end{figure}

For $\rho_x = \rho_y \equiv \rho$,
the cross-polar level and the differential response are
\begin{equation}
 \begin{split}
  X&P(\theta)
   = \frac{1}{4} \frac{\sin^2 k_z d}{\sin^2 k d}
  \cos^2 \theta \left( 1 - \cos \theta \right)^2
  \\
  & \times
  \frac{\left| \eta \sin kd - i \rho e^{i kd} \right|^2
  \left| \eta \sin k_z d + i \rho e^{i k_z d} \right|^2
  }{
  \left| \eta \sin k_z d \cos \theta - i \rho e^{i k_z d} \right|^2
  \left| \eta \sin k_z d - i \rho e^{i k_z d}  \cos \theta \right|^2
  },
 \end{split}
 \label{equ:XP_w_backshort}
\end{equation}
\begin{equation}
 \begin{split}
  D&R(\theta)
  = \frac{1}{8} \frac{\sin^2 k_z d}{\sin^2 kd} \sin^2 2\theta
  \\
  & \times
  \frac{\left|\eta^2 \sin^2 k_z d - \rho^2\right|
  \left| \eta \sin kd - i \rho e^{i kd} \right|^2
  }{
  \left| \eta \sin k_z d \cos \theta - i \rho e^{i k_z d} \right|^2
  \left| \eta \sin k_z d - i \rho e^{i k_z d}  \cos \theta \right|^2
  }.
 \end{split}
 \label{equ:DR_w_backshort}
\end{equation}
Note that Eqs.~(\ref{equ:XP_w_backshort}) and (\ref{equ:DR_w_backshort})
are equivalent to the no-backshort case [Eqs.~(\ref{equ:xp_no_bs}) and
(\ref{equ:dr_no_bs})] only if one substitutes
$k_z d = k d = \pi/2$ and $\rho \rightarrow 2\rho$.
Figure~\ref{fig:xp_dr_with_bs} shows $XP(\theta)$ and $DR(\theta)$
for $\rho=\eta$ and various $d$.  For
$d \sim 1.4 \lambda/4$, a low
level of the beam systematics, $XP$ below $-30$\,dB and $DR$ below
$-20$\,dB, can be achieved for a wide range of acceptance angle,
$\theta \lesssim 55^\circ$.
The $XP$ and $DR$ do not significantly degrade for $d = 1.2 \lambda/4$
or $1.6 \lambda/4$, suggesting a fractional bandwidth of $\sim 30\,\%$ is possible
while maintaining the low levels of systematics.

\begin{figure}[tbp]
 \includegraphics[width=0.45\textwidth]{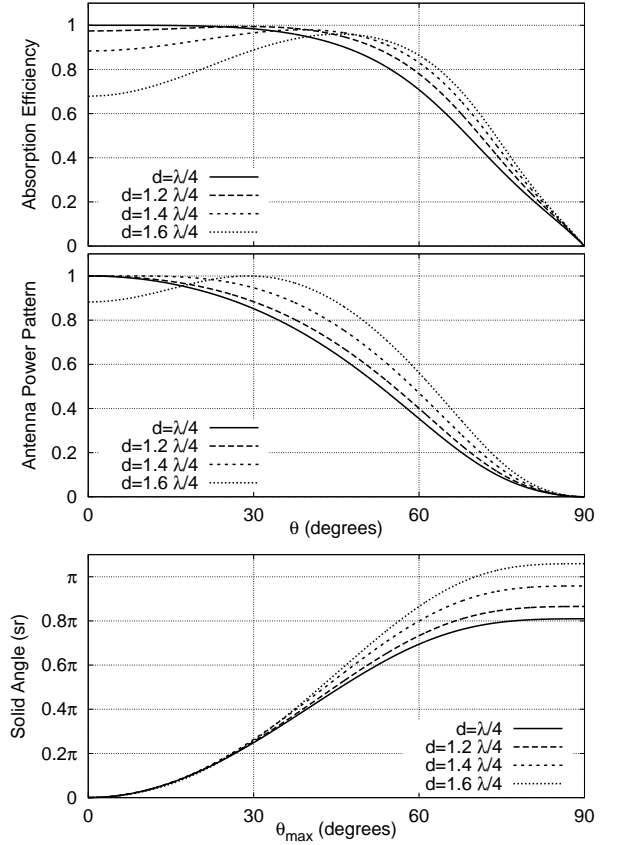}
 \caption{\label{fig:efficiency_app_with_backshort} 
 Absorption efficiency (top), antenna power pattern (middle), and the
 solid angle
 (bottom) for with $\rho_x = \rho_y = \eta$ and with a
 reflective backshort.  Cases with the backshort positions of
 $d = \lambda/4$, $d = 1.2 \lambda/4$, $d = 1.4 \lambda/4$, and
 $d = 1.6 \lambda/4$ are shown.
 }
\end{figure}

We define the angular dependence of the
power absorption per unit area of the resistive surface, $P_a$, 
as an average over $\phi$:
\begin{equation}
 P_a(\theta) \equiv \frac{1}{2\pi} \int_{0}^{2\pi} \!\!\!\!
  \mathrm{d}\phi \, P_a(\theta, \phi) \:.
\end{equation}
For one grid running in $x$ direction,
$P_a(\theta, \phi)$ here is
\begin{equation}
 \begin{split}
 P_a(\theta, \phi)
  & = \rho_x {j_x}^2
  \\
  & =
  \frac{1}{\eta}
  \left< 
  \left| \widetilde{G}_{CO}^{v}(\theta,\phi) a_v + 
         \widetilde{G}_{XP}^{v}(\theta,\phi) a_h \right|^2 
   \right>
  \\
  & = 
  \frac{1}{\eta}
  \left( \left| \widetilde{G}_{CO}^{v}(\theta,\phi) \right|^2 
  + \left| \widetilde{G}_{XP}^{v}(\theta,\phi) \right|^2 \right)
  a^2 \:,
 \end{split}
\end{equation}
where the last equality assumes unpolarized incident light with
$\left< |a_v|^2 \right> = \left< |a_h|^2 \right> = a^2$
and $\left< a_v^* a_h \right> = 0$.
When $\rho_x = \rho_y = \eta$,
$P_a(\theta)$ is
\begin{equation}
 \begin{split}
  P_a(\theta) & = \sin^2 k_z d \, \cos^2 \theta \: \eta^{-1} a^2
  \\
  &
  \times
  \left[4 \sin^2 k_z d \, (1+\cos\theta)^2 + \cos^2 k_z d \, (3+\cos
  2\theta)\right]
  \\
  & \times \frac{1}
  { \sin^2 k_z d \, (1+\cos\theta)^2 + \cos^2 k_z d }
  \\
  & \times \frac{1}
  {\sin^2 k_z d \, (1+\cos\theta)^2 + \cos^2 k_z d \, \cos^2 \theta }
  \:.
 \end{split}
 \label{equ:antenna_power_with_backshort}
\end{equation}
Again, Eq.~(\ref{equ:antenna_power_with_backshort}) is equivalent to
Eq.~(\ref{equ:power_absorption_no_bs}) if one formally substitutes
$k_z d = \pi/2$, except
for an overall factor of two.
Substituting Eq.~(\ref{equ:antenna_power_with_backshort}) into 
the first equalities of Eqs.~(\ref{equ:efficiency_def}) and
(\ref{equ:antenna_power_pattern_def}),
we define the absorption efficiency $\epsilon(\theta)$ and the antenna
power pattern $P(\theta)$, 
respectively.  Similarly, we define the solid angle $\Omega (\theta_{\max})$ by the first
equality of Eq.~(\ref{equ:solid_angle_wo_bs}).
Figure~\ref{fig:efficiency_app_with_backshort} shows them for various
backshort position $d$ for $\rho_x = \rho_y = \eta$.
Unlike the case without a backshort,
$\theta = 0$ does not always give the maximum $P_a(\theta)$;
e.g., $P_a(0) \neq \max_\theta P_a(\theta)$
when $1.4 \lambda/4 \lesssim d < \lambda/2$.

In summary,
the use of the reflective backshort leads to significant improvement of
the absorption efficiency and maintains a large sensor acceptance angle.
On the other hand, it also leads to non-zero cross polarization and
differential response.  However, 
these systematics can be suppressed
 for a  wide range of incident
angles, $0\leq \theta \lesssim 60^\circ$,
by matching resistivity to the free-space impedance
and optimizing the backshort position.
We also point out
that Eq.~(\ref{equ:alpha_beta_with_backshort})
shows that the differential response (sourced by
the $\delta^\mathrm{bs}$ term in $\alpha_\pm^\mathrm{bs}$)
 depends on $\phi$ as $\cos 2\phi$,
yielding an spurious polarization pattern with even
parity only and maintaining a systematic-free measurement of the odd-parity
patterns~\cite{2003PhRvD..67d3004H,2008PhRvD..77h3003S}.
This is of importance
in the context of CMB polarization observation,
where the precision measurement of the so-called $B$-mode, or parity odd, pattern is the primary
goal.

In order to see why the beam systematics are small 
and yet what prevents their exact elimination,
we look into the beam patterns in the $E$ and $H$ planes.
For $\phi=0$,
$\widetilde{G}_{XP}^h(\theta,0) = \widetilde{G}_{XP}^v(\theta,0) = 0$.
When $\rho_x = \rho_y = \rho$, the coupling coefficients are
\begin{equation}
 \begin{split}
  \widetilde{G}_{CO}^v(\theta,0) & 
  = \frac{e^{i k_z d } 2 \rho \cos \theta}{\eta \cos \theta - i \rho
  e^{- i k_z d} \left( \sin k_z d \right)^{-1}} \:, \\
  \widetilde{G}_{CO}^h(\theta,0) & 
  = \frac{e^{i k_z d } 2 \rho \cos \theta}{\eta - i \rho e^{i k_z d}
  \cos \theta
  \left( \sin k_z d \right)^{-1}} \:.
 \end{split}
\end{equation}
We reform them and relate the incident field and the induced current
using Eq.~(\ref{equ:G__ampl_vs_current}) as:
\begin{equation}
 \begin{split}
  a_v & = \frac{e^{- i k_z d}}{2 \cos \theta}\left(\eta \cos \theta
  + \frac{\rho }{1 - e^{-i k_z d} \cos k_z d}\right) j_x
  \\
  a_h & = \frac{e^{- i k_z d}}{2 \cos \theta}\left(\eta
  +  \frac{\rho \cos \theta}{1 - e^{-i k_z d} \cos k_z d}\right) j_y \:,
 \end{split}
 \label{equ:h_vs_e_plane_current_vs_incident_w_bs}
\end{equation}
where $a_v$ ($a_h$) is the electric-field amplitude for $E$-plane
($H$-plane) incident since $\phi=0$.  

Equation~(\ref{equ:h_vs_e_plane_current_vs_incident_w_bs}) has an almost
equivalent 
form to the no-backshort case of
Eq.~(\ref{equ:h_vs_e_plane_current_vs_incident})
 except for an extra factor of $-e^{-i k_z d} \cos k_z d$
in the second term; without this factor,
the beam
pattern is symmetric between $E$ and $H$ planes when
$\rho = \eta$.  As we see below, this extra
factor can be understood as a reception pattern of a two-element
phased-array antenna.  Note that the overall phase
of $e^{- i k_z d}$
in Eq.~(\ref{equ:h_vs_e_plane_current_vs_incident_w_bs})
is physically irrelevant, being due to our
convention here that the resistive sheet is placed at $z=-d$, not at
$z=0$.

\begin{figure}[tbp]
 \begin{center}
 \includegraphics[width=0.47\textwidth]{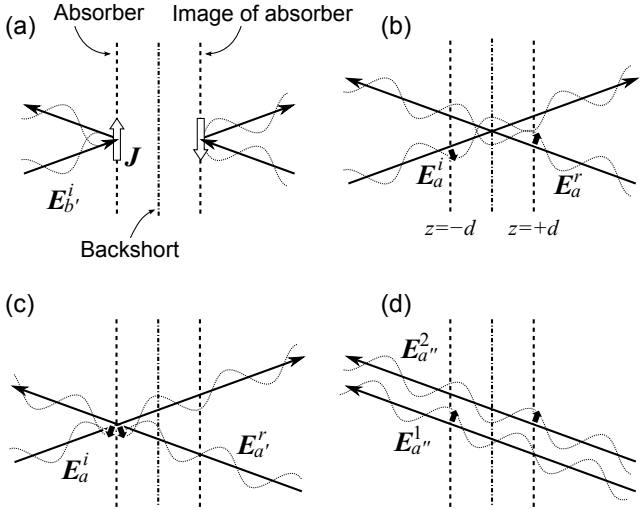}
  \caption{ \label{fig:intuitive_understanding_with_bs} 
  Schematic figures of the field
  interacting with an electrically thin absorber and a backshort,
  (a) at the limit of $\rho \ll \eta$
  and (b) at the limit of $\rho \gg \eta$.
  The latter is further decomposed into two: (c) where the
  total field is the
  sum of the incident field $\vect{E}_{a}^i$
  and a component of the reflected field $\vect{E}_{a'}^r$
  that is phase coherent to the incident field on the surface of
  $z=-d$, 
  and (d) where the field is a superposition of
  $\vect{E}_{a''}^1$ ($\equiv - \vect{E}_{a'}^r$)
  and $\vect{E}_{a''}^2$ ($\equiv \vect{E}_{a}^{r}$).
  }
 \end{center}
\end{figure}

We again look into the terms in
Eq.~(\ref{equ:h_vs_e_plane_current_vs_incident_w_bs}) by considering the
limits of $\rho \ll \eta$ (approximately shorted) and
 $\rho \gg \eta$ (approximately open).
Figure~\ref{fig:intuitive_understanding_with_bs} illustrates these limits.
The first term, which dominates in the limit of
$\rho \ll \eta$, corresponds to a reflection off a perfect
conductor sheet (Fig.~\ref{fig:intuitive_understanding_with_bs}a).
In this case, the mirror image in the region $z>0$ has no effect
on the sheet located at $z=-d$ and thus
Eq.~(\ref{equ:electromotive_force}) describes this 
component in the same manner as in the case without a backshort
(cf. Fig.~\ref{fig:boundary_condition}c).

Slightly more complicated are the second terms
in Eq.~(\ref{equ:h_vs_e_plane_current_vs_incident_w_bs}), which
dominate when
$\rho \gg \eta$.
These correspond to the incident field, $\vect{E}_a^i$,
and its mirror conjugate,
which corresponds to the reflected field $\vect{E}_a^r$,
 passing through the absorber with negligible
interactions (Fig.~\ref{fig:intuitive_understanding_with_bs}b).
It is convenient to decompose $\vect{E}_a^r$
into two parts, one of which, called $\vect{E}_{a'}^{r}$,
is phase coherent with $\vect{E}_{a}^i$ at $z=-d$
and satisfies
$\left|\vect{E}_{a'}^{r}\right| = \left|\vect{E}_a^r\right|
=\left|\vect{E}_a^i\right|$
(Fig.~\ref{fig:intuitive_understanding_with_bs}c and d).
Then the entire field is represented with 
two combinations:  (c) $\vect{E}_{a}^i$
and $\vect{E}_{a'}^{r}$;
and (d)
a superposition of $\vect{E}_{a''}^1$
($\equiv - \vect{E}_{a'}^r$)
and $\vect{E}_{a''}^2$ ($\equiv \vect{E}_{a}^{r}$).
Note that $\vect{E}_{a''}^1\bigr|_{z=-d}$ has the same phase
as $\vect{E}_{a''}^2\bigr|_{z=+d}$ by construction, since
the image theory forces the field
$\vect{E}_{a}^{r}\bigr|_{z=+d}$ to have a phase shifted by  $\pi$
compared to that of $\vect{E}_{a}^i\bigr|_{z=-d}$.
The system with $\rho=\eta$ would recover the symmetry between $E$ and $H$ planes,
or the symmetry between Eqs~(\ref{equ:voltage_drop_due_to_resistivity})
and (\ref{equ:electromotive_force}),
if there were contributions only from 
(a)
and (c).
The relation of the field and current in (c) is the same
as that of
Eq.~(\ref{equ:voltage_drop_due_to_resistivity}), 
except the current here is twice as large since $\vect{E}_{a'}^{r}$ is phase
coherent to $\vect{E}_{a}^i$ at $z=-d$:
\begin{equation}
  \rho \left| \vect{J}_{a'} \right|
    = \cos \theta_E \left( \left| \vect{E}_{a}^i \right|
  + \left| \vect{E}_{a'}^r \right|
  \right)
     = 2 \cos \theta_E \left| \vect{E}_i \right| \:.
  \label{equ:voltage_drop_due_to_resistivity_w_bs}
\end{equation}
The contribution from (d) introduces a small violation of the symmetry.
This component resembles the
reception
pattern of a two-element phased-array antenna with uniform (or
zero-phase)
excitation:
\begin{equation}
   \rho \left| \vect{J}_{a''} \right|
    = \cos \theta_E \left| \vect{E}_{a'}^1 + \vect{E}_{a'}^2 \right|
    = 2 \cos \theta_E \left| \cos k_z d  \right| \left| \vect{E}_i \right| \:,
 \label{equ:phased_array_current_field}
\end{equation}
where $\cos k_z d = \cos \left( k d \cos \theta \right)$ is the space
factor of the phased array.
The relation of the current and the incident field in the limit of
$\rho\gg\eta$ (Fig.~\ref{fig:intuitive_understanding_with_bs}b)
corresponds to the sum of
Eqs.~(\ref{equ:voltage_drop_due_to_resistivity_w_bs})
and (\ref{equ:phased_array_current_field}):
\begin{equation}
 \rho \left| \vect{J} \right|
  = 2 \cos \theta_E \left| 1 - e^{-i k_z d} \cos k_z d \right| \left| \vect{E}_i \right|
  \:,
 \label{equ:voltage_drop_with_bs}
\end{equation}
where the phase factor $-e^{-i k_z d}$ arises as a phase
difference between (c) and (d);
the induced current in (c) is in-phase with $\vect{E}_i$
at $z=-d$ while that in (d) is in-phase with $\vect{E}_i$ at $z=0$.

When the surface resistivity is at the neither limits of
$\rho \ll \eta$ or $\rho \gg \eta$,
the incident electric field corresponds to the sum of those in
Eqs.~(\ref{equ:electromotive_force})
and (\ref{equ:voltage_drop_with_bs}):
\begin{equation}
 2 \left| \vect{E}_i \right|
  =
  \left| \frac{\eta}{\cos \theta_H}
   + \frac{\rho}{\cos \theta_E} 
   \frac{1}{1 - e^{-i k_z d} \cos k_z d}
  \right| \left| J \right| \:,
  \label{equ:intuitive_boundary_condition_wth_bs}
\end{equation}
which represents the same boundary condition as that discussed
in Fig.~\ref{fig:boundary_condition} for
Eq.~(\ref{equ:j_vs_e_h_surface}).
Equation~(\ref{equ:intuitive_boundary_condition_wth_bs}) 
is equivalent to
Eq.~(\ref{equ:h_vs_e_plane_current_vs_incident_w_bs})
obtained from the exact calculation done in
Appendix~\ref{sec:appendix_calc_with_backshort}.

When $\rho = \eta$, the $E$- and $H$-plane responses are maximally
symmetric and the residual arises from the phased-array type
contribution of
Eq.~(\ref{equ:phased_array_current_field}).
More strictly, the systematics 
arises due to the difference of this
component between $E$ and $H$ planes;
the difference of the coefficient
in Eq.~(\ref{equ:phased_array_current_field})
is
$[1-\cos \theta] \cos k_z d$ since $\theta_E = \theta$  
($\theta_E = 0$) for $E$-plane ($H$-plane) incident.
This is consistent with the fact that
$\delta^{\mathrm{bs}}(\rho=\eta,d,\theta) \propto [1-\cos \theta] \cos k_z d$ as
in Eq.~(\ref{equ:delta_with_backshort}).

We close the discussion here by pointing out that a reflective backshort
made of
perfect magnetic conductor (PMC), as opposed to perfect electric
conductor (PEC) or metal,
 would eliminate the polarization systematics.  With
 the PMC backshort, the ideal configuration of the 
 absorber sheet and the backshort is to place them infinitesimally close to
 each other.  In this case, the phased-array type contribution vanishes and
 the symmetry between $E$ and $H$ planes is recovered.
Such a configuration is indeed mentioned in
the current sheet model as an implementation of an ideal current
sheet~\cite{1965ITAP...13..506W,hansen2009phased}.
However, a practical implementation of a PMC tends to have restrictions
such as finite bandwidth.  This is because an implementation of PMC
often consists of a PEC and a quarter wave delay, and the latter
introduces a wavelength dependence.  Indeed, one can regard the PEC
backshort discussed above as a PMC backshort right behind the absorber,
where the spatial distance $d$ is the quarter wave delay.  A phase delay
realized by the spatial separation depends both on the incident angle
and on the frequency, and only exactly realizes the desired boundary at
a single frequency.  Thus, it leads to the systematics discussed above in
detail.

\section{Diffraction Effect due to Finite Size of the Absorber}
\label{sec:diffraction_consideration}
In this section we argue that diffraction effects due to the finite size of the absorber are minor.
Such diffraction effects are discussed in detail in
Refs.~\cite{2003SPIE.4855...49W,2008PASP..120..430C,Thomas:10}.
Reference~\cite{2008PASP..120..430C}
shows that the net differential response can be approximated as
$\sim (\lambda/p) \cdot 9$\,\%,
where $p$ is the size of the absorber.
As expected, the larger $p$ is, the smaller the differential
response.
That analysis also shows that the differential response has
a $\cos 2\phi$ ($\sin 2\phi$) pattern for
a channel sensitive to $Q$ ($U$) polarization.
This is significant for the purpose of CMB polarization
measurement, as 
the spurious polarization sourced by this effect
does not create the so-called $B$-mode, or parity odd,
pattern~\cite{2003PhRvD..67d3004H,2008PhRvD..77h3003S}.

The analysis in Ref.~\cite{2008PASP..120..430C} is for a blackbody
absorber or radiator.
However, the difference between an electrically thin resistive absorber of
interest
 and a blackbody absorber is
 minor, and only involves angular response pattern, where the
blackbody absorber follows $\cos \theta$
while the absorber sheet follows a different functional shape,
Eq.~(\ref{equ:antenna_power_pattern_def}), for example.
Their difference is small (see, e.g., Fig.~\ref{fig:beam_shape}),
and is only
significant for large angle incident waves, $\theta>60^\circ$, which
would either be
stopped by baffling or have little contribution to the total solid angle.
Thus, the diffraction estimate for the blackbody absorber
 serves as a good order-of-magnitude estimate for an electrically thin resistive
 absorber, too.
A more rigorous estimate of the thin absorber response is possible with 
the extension presented in Ref.~\cite{Thomas:10}.

We also note that these analyses adopt metal boundary conditions,
taking the frame around the absorber as a perfect conductor.
Strictly, the diffraction effect depends on the boundary condition and
the metal boundary is one of the worst cases.  The diffraction pattern
in general depends on the impedance of the aperture
frame~\cite{1996ITAP...44.1464Y}.

There are other possible effects that arise due to the finite size of the
absorber.  For example, the edge of the absorber, which may have
different impedance from the main absorbing area, could affect the image
quality.
Effects of this type are dependent on specific implementations of
devices and
the configurations in which the devices are placed,
such as edge tapering of the absorber resistivity and baffling around
the absorber surface.
These implementation dependent effects are beyond the scope of this
paper.

\section{Conclusion}
Estimates of polarization beam patterns 
are presented for a multimoded bolometer employing an electrically thin resistive
absorber at the limit of infinitely large area ($N_\mathrm{modes} \gg 1$).
For a freestanding
thin absorber, cross polarization and differential response
can be eliminated by choosing the surface resistivity to be half of
the free-space impedance [Eqs.~(\ref{equ:xp_no_bs}) and (\ref{equ:dr_no_bs})].
The absorption efficiency can be significantly improved by employing a reflective
backshort termination.  Although the backshort introduces non-zero
cross polarization and differential response,
the levels of these systematics can be suppressed by choosing the
resistivity matched to the free-space impedance and optimizing the
position of the backshort.
For a practical application where the absorbing area is finite,
diffraction sources additional beam systematics.  However,
the differential response due to the diffraction is small when the
number of modes is large.

The low levels of cross polarization lead to high efficiency
in polarization detection.
The smallness of the differential response is critical for an accurate
measurement
of polarization, as it can create spurious polarization from 
fluctuations in intensity.
One significant potential uses for these sensors is to measure the
signature of inflation during the early epic of the universe,
which manifests itself as odd-parity patterns in the CMB
polarization~\cite{1997PhRvL..78.2054S,1997PhRvL..78.2058K}.
The differential responses due to the two types of residual
systematics mentioned above, one due to a backshort and another due to
diffraction,
have  angular patterns of even parity
and do not contaminate the inflation signature~\cite{2003PhRvD..67d3004H,2008PhRvD..77h3003S}.

\section*{Acknowledgment}
This work was done in the context of developing experimental projects
using multimoded detectors, in particular MuSE and PIXIE.  We thank the
collaborators of these projects; especially we acknowledge D.~J.~Fixen,
A.~J.~Kogut, S.~S.~Meyer, and S.~T.~Staggs for their
insights and persistent encouragement.
We thank L.~A.~Page and N.~Jarosik for fruitful discussions.
A.~K. acknowledges the Dicke Fellowship.

\appendix

\section{Explicit Definitions of Unit Vectors
 \label{sec:vect_def}
 }
Unit vectors that define Cartesian coordinates are:
\begin{equation}
 \ex \equiv \left(
	     \begin{array}{c}
	      1\\
	      0\\
	      0\\
	     \end{array}
	    \right) \:,
 \quad
 \ey \equiv \left(
	     \begin{array}{c}
	      0\\
	      1\\
	      0\\
	     \end{array}
	    \right) \:,
 \quad
 \ez \equiv \left(
	     \begin{array}{c}
	      0\\
	      0\\
	      1\\
	     \end{array}
	    \right) \:.
\end{equation}
Unit vectors for polar coordinates are:
\begin{equation}
 \begin{split}
 \er \equiv \left(
	     \begin{array}{c}
	      \sin \theta \cos \phi \\
	      \sin \theta \sin \phi \\
	      \cos \theta \\
	     \end{array}
  \right) & \:, \quad
 \eq \equiv \left(
	     \begin{array}{c}
	      \cos \theta \cos \phi \\
	      \cos \theta \sin \phi \\
	      -\sin \theta \\
	     \end{array}
	    \right) \:,
  \\
 \ep \equiv & \left(
	     \begin{array}{c}
	      -\sin \phi \\
	      \cos \phi \\
	      0\\
	     \end{array}
	    \right) \:.
 \end{split}
\end{equation}
Here are unit vectors that describe the fields
of the two polarization bases in Ludwig's third
definition~\cite{1973ITAP...21..116L}:
\begin{equation}
 \begin{split}
 \ev
  & \equiv
  \eq \cos \phi - \ep \sin \phi
  \\
  & =
  \left(
   \begin{array}{c}
    \cos \theta \cos^2 \phi + \sin^2 \phi \\
    -(1-\cos \theta) \sin \phi \cos \phi \\
    - \sin \theta \cos \phi \\
   \end{array}
  \right) \:,
 \end{split}
\end{equation}
\begin{equation}
 \begin{split}
 \eh
  & \equiv
  \eq \sin \phi + \ep \cos \phi
  \\
  & =
  \left(
   \begin{array}{c}
    -(1-\cos \theta) \sin \phi \cos \phi \\
    \cos \theta \sin^2 \phi + \cos^2 \phi \\
    - \sin \theta \sin \phi \\
   \end{array}
  \right) \:.
 \end{split}
\end{equation}
They are perpendicular to $\er$ by construction and satisfy
\begin{equation}
 \er \times \ev = \eh \quad \mathrm{and} \quad \er \times \eh = -\ev \:.
\end{equation}
Parallel and perpendicular polarization vectors for $x$-$z$ plane
incident waves are:
\begin{equation}
 \eparal
  \equiv
  \left(
   \begin{array}{c}
    \cos \theta \\
    0 \\
    -\sin \theta \\
   \end{array}
  \right) \:, \quad
 \eperp\equiv
  \left(
   \begin{array}{c}
    0 \\
    1 \\
    0 \\
   \end{array}
  \right) \:.
\end{equation}
They are related to ($\eq$, $\ep$) and ($\ev$, $\eh$) by
\begin{equation}
 \begin{split}
 \vect{R}_z(\phi) \eparal
  & = \eq = \cos \phi \ev + \sin \phi \eh \:,
  \\
 \vect{R}_z(\phi) \eperp
  & = \ep = -\sin \phi \ev + \cos \phi \eh \:,
 \end{split}
 \label{equ:perp_paral_theta_phi}
\end{equation}
where $\vect{R}_z(\phi)$ is a rotation matrix around the $z$ axis:
\begin{equation}
 \vect{R}_z(\phi)
  \equiv
  \left(
   \begin{array}{ccc}
    \cos \phi & -\sin \phi & 0 \\
    \sin \phi & \cos \phi & 0 \\
    0 & 0 & 1 \\
   \end{array}
  \right) \:.
\end{equation}

\section{Analytic calculation of the response of freestanding thin
 absorber}
 \label{sec:appendix_calc_without_backshort}
We consider the arrangement shown in Fig.~\ref{fig:no_backshort}.
The electric and magnetic fields of an arbitrary polarized plane wave
can be expressed as a linear combination of 
vertically and holizontally polarized components with amplitudes
$a_v \equiv a(\theta, \phi, v)$ and
 $a_h \equiv a(\theta, \phi, h)$:
\begin{equation}
 \begin{split}
 \vect{E}_i(\vect{r};\vect{k})
  & = e^{-i \vect{k}\cdot \vect{r}}
  \left(
   a_v \ev + a_h \eh
  \right) \\
  & =
  e^{-i \vect{k}\cdot \vect{r}}
   \vect{R}_z(\phi)
   \left(
    a_\parallel \eparal  + a_\perp \eperp
   \right) \:,
  \\
 \vect{B}_i(\vect{r};\vect{k})
  & = \frac{1}{c} e^{-i \vect{k}\cdot \vect{r}}
  \left(
   a_v \eh - a_h \ev
  \right) \\
  & =
  \frac{1}{c} e^{-i \vect{k}\cdot \vect{r}}
   \vect{R}_z(\phi)
   \left(
    a_\parallel \eperp - a_\perp \eparal
   \right)
  \:.
 \end{split}
 \label{equ:wave_incident}
\end{equation}
Amplitudes $a_\parallel$
and $a_\perp$ are defined for convenience and related to 
$a_v$ and $a_h$ as
\begin{equation}
 \left(
  \begin{array}{c}
   a_v \\
   a_h \\
  \end{array}
 \right)
 =
 \vect{R}
 \left(
  \begin{array}{c}
   a_\parallel \\
   a_\perp \\
  \end{array}
 \right) \:,
 \label{equ:vert_hor_to_paral_perp}
\end{equation}
with
\begin{equation}
 \vect{R}
  \equiv 
  \left(
   \begin{array}{ccc}
    \cos \phi & -\sin \phi \\
    \sin \phi & \cos \phi \\
   \end{array}
  \right) \:.
\end{equation}
The reflected and transmitted fields
can similarly be written as 
\begin{equation}
 \begin{split}
 \vect{E}_r(\vect{r};\vect{k})
  & = e^{-i \vect{\bar{k}}\cdot \vect{r}}
   \vect{R}_z(\phi+\pi)
  \left(
   b_\parallel \eparal + b_\perp \eperp
  \right) \:,
  \\
 \vect{B}_r(\vect{r};\vect{k})
  & = -\frac{1}{c} e^{-i \vect{\bar{k}}\cdot \vect{r}}
   \vect{R}_z(\phi+\pi)
  \left(
   b_\parallel \eperp - b_\perp \eparal
  \right) \:,
 \end{split}
 \label{equ:wave_refl}
\end{equation}
and
\begin{equation}
 \begin{split}
  \vect{E}_t(\vect{r};\vect{k})
  & = e^{-i \vect{k}\cdot \vect{r}}
   \vect{R}_z(\phi)
  \left(
   c_\parallel \eparal + c_\perp \eperp
  \right) \:,
  \\
 \vect{B}_t(\vect{r};\vect{k})
  & = \frac{1}{c} e^{-i \vect{k}\cdot \vect{r}}
   \vect{R}_z(\phi)
  \left(
   c_\parallel \eperp - c_\perp \eparal
  \right) \:,
 \end{split}
 \label{equ:wave_transm}
\end{equation}
respectively, where $\vect{\bar{k}} \equiv (k_x,k_y,-k_z)^T$; and
$b_{\parallel, \perp}$ and $c_{\parallel, \perp}$ are the amplitudes of
the transmitted and reflected waves, respectively.

The absorber can either be a membrane or a pair of orthogonal resistive
wire grids
running in the $x$ and $y$ directions.
For the latter, we allow each grid
to have different resistivity, including the case
where one of the grids is absent (i.e., infinite resistivity).
At $z=0$, where the absorber is, all three plane wave components
have the phase
$e^{-i(k_x x +k_y y)}$ and thus the surface current density on
the absorber induced by the incident field can be
written as
\begin{equation}
 \vect{J}(x,y,z) = \delta(z) e^{-i(k_x x +k_y y)}
  \left( j_x \ex + j_y \ey \right) \:,
\end{equation}
where $j_x$ and $j_y$ are complex current amplitudes.

The vector
potential $\vect{A}$ of the field induced by the current $\vect{J}$
is the solution of the following Helmholtz
equation:
\begin{equation}
 \nabla^2 \vect{A} + k^2 \vect{A} = - \mu \vect{J} \:,
\end{equation}
with magnetic permeability $\mu$.  Note that 
we implicitly choose the Lorenz gauge by adopting the Helmholtz equation.
The solution is
\begin{equation}
 \vect{A} = \frac{\mu}{2ik_z} e^{-i k_x x -i k_y y -i k_z
  \left|z\right|}
  \left( j_x \ex + j_y \ey \right) \:,
\end{equation}
and the corresponding electric and magnetic fields ($z\neq 0$) are
\begin{equation}
 \begin{split}
 \vect{E}(\vect{r};\vect{k})
  & = \frac{c}{i k} \nabla \times (\nabla \times \vect{A}) \:,
  \\
 \vect{B}(\vect{r};\vect{k})
  & = \nabla \times \vect{A} \:.
 \end{split}
\end{equation}
Thus, the field discontinuity at $z=0$ due to the
current $\vect{J}$ is
\begin{equation}
 \begin{split}
 \vect{E} \Bigr|_{z \rightarrow -0} & - \vect{E} \Bigr|_{z
  \rightarrow +0}
  \\
  & =
  \eta e^{-i k_x x -i k_y y}
  \sin \theta \left(j_x  \cos \phi + j_y  \sin \phi \right)
  \ez \:,
  \\
 \vect{B} \Bigr|_{z \rightarrow -0} & - \vect{B} \Bigr|_{z
  \rightarrow +0}
  \\
  & =
  \mu e^{-i k_x x -i k_y y}
  \left( - j_y \ex + j_x \ey \right) \:,
 \end{split}
 \label{equ:discontinuity}
\end{equation}
where $\eta\equiv \sqrt{\mu/\epsilon}$ is the wave impedance 
the surrounding medium (e.g., vacuum).
The electric and magnetic fields
of the left hand side
 are related to 
Eqs.~(\ref{equ:wave_incident}), (\ref{equ:wave_refl}), and
(\ref{equ:wave_transm}) evaluated at $z=0$.
This leads
 to the following boundary conditions
on the field amplitudes:
\begin{equation}
 \begin{split}
  a_\parallel - b_\parallel - c_\parallel & = 0  \:,
  \\
  a_\perp - b_\perp - c_\perp & = 0 \:,
 \end{split}
 \label{equ:boundary1}
\end{equation}
and
\begin{equation}
 \begin{split}
  a_\parallel + b_\parallel - c_\parallel
  & =  \eta \left( j_x \cos \phi  +  j_y \sin \phi \right) \:,
  \\
  a_\perp + b_\perp - c_\perp
  & =  - \frac{\eta}{\cos \theta} \left( j_x \sin \phi - j_y \cos \phi \right) \:,
 \end{split}
 \label{equ:boundary2}
\end{equation}
where the former [latter] comes from the electric [magnetic] field
component of Eq.~(\ref{equ:discontinuity}).

In addition to the Maxwell equations used above, the current density and
the electric fields at $z=0$ are related by Ohm's law, leading to
\begin{equation}
 \left(
  \begin{array}{c}
   \rho_x j_x   \\
   \rho_y j_y   \\
  \end{array}
 \right)
 =
 \vect{R}
 \left(
  \begin{array}{c}
   c_\parallel \cos \theta \\
   c_\perp  \\
  \end{array}
 \right) \:,
 \label{equ:ohmic_law}
\end{equation}
where $\rho_x$ and $\rho_y$ are the surface resistivities
of the absorber in the $x$ and $y$ directions, respectively.

Solving Eqs.~(\ref{equ:boundary1}), (\ref{equ:boundary2}) and (\ref{equ:ohmic_law}),
we obtain
\begin{equation}
 \left(
  \begin{array}{c}
   c_\parallel \\
   c_\perp \\
  \end{array}
 \right)
 =
 \vect{\mathcal{T}}
 \left(
  \begin{array}{c}
   a_\parallel \\
   a_\perp \\
  \end{array}
 \right) ,
\end{equation}
\begin{equation}
 \left(
  \begin{array}{c}
   b_\parallel \\
   b_\perp \\
  \end{array}
 \right)
 =
 \vect{\mathcal{R}}
 \left(
  \begin{array}{c}
   a_\parallel \\
   a_\perp \\
  \end{array}
 \right) ,
\end{equation}
\begin{equation}
 \left(
  \begin{array}{c}
   j_x \\
   j_y \\
  \end{array}
 \right)
 =
 \vect{\mathcal{A}}
 \left(
  \begin{array}{c}
   a_\parallel \\
   a_\perp \\
  \end{array}
 \right) ,
  \label{equ:absorption_in_perp_parall}
\end{equation}
where $\vect{\mathcal{T}}$, $\vect{\mathcal{R}}$,
and $\vect{\mathcal{A}}$ are transmission, reflection, and absorption
matrices defined as
\begin{equation}
 \vect{\mathcal{T}}
  \equiv 
  \left[
   \vect{1}
   +
   \frac{\sec \theta}{2}
   \vect{Q} \vect{R}^{-1} \vect{G} \vect{R} \vect{Q}
  \right]^{-1} ,
\end{equation}
\begin{equation}
 \begin{split}
 \vect{\mathcal{R}}
  & \equiv \frac{1}{2} \sec \theta
  \vect{Q} \vect{R}^{-1} \vect{G} \vect{R} \vect{Q} \vect{\mathcal{T}}
  \\
  &
  =
  \left[
  \vect{1} + 2 \cos \theta
  \left(
  \vect{Q} \vect{R}^{-1} \vect{G} \vect{R} \vect{Q}
  \right)^{-1}
  \right]^{-1} \:,
 \end{split}
\end{equation}
\begin{equation}
 \vect{\mathcal{A}}
  \equiv
  \eta^{-1}
  \left[
   \vect{Q}^{-1} \vect{R}^{-1} \vect{G}^{-1}
   +
   \frac{\sec \theta}{2}
    \vect{Q} \vect{R}^{-1}
  \right]^{-1} ,
\end{equation}
with
\begin{equation}
 \vect{Q} \equiv
  \left(
   \begin{array}{cc}
    \cos \theta & 0 \\
    0 & 1 \\
   \end{array}
  \right) , \quad 
  \vect{G}
  \equiv
  \left(
   \begin{array}{cc}
    \eta / \rho_x & 0 \\
    0 & \eta / \rho_y \\
   \end{array}
  \right) .
\end{equation}
Combining Eqs.~(\ref{equ:vert_hor_to_paral_perp}) and
(\ref{equ:absorption_in_perp_parall}), we obtain
\begin{equation}
 \left(
  \begin{array}{c}
   j_x \\
   j_y \\
  \end{array}
 \right)
 =
 \vect{\mathcal{A}} \vect{R}^{-1}
 \left(
  \begin{array}{c}
   a_v \\
   a_h \\
  \end{array}
 \right)
 \equiv
 \vect{\Sigma}
 \left(
  \begin{array}{c}
   a_v \\
   a_h \\
  \end{array}
 \right), 
 \label{equ:j_vs_sigma_vs_ampl}
\end{equation}
where
\begin{equation}
 \vect{\Sigma}^{-1}
  =
  \frac{1}{4 \cos \theta}
 \left(
  \begin{array}{cc}
   \alpha_+ (\rho_x) & \beta (\rho_y) \\
   \beta (\rho_x) & \alpha_- (\rho_y) \\
  \end{array}
 \right) \:,
 \label{equ:sigma_def}
\end{equation}
with
\begin{equation}
 \begin{split}
 \alpha_\pm (\rho)
  & \equiv
  (2 \rho + \eta) (1 + \cos \theta) \pm
  \cos 2 \phi (2 \rho - \eta) (1 - \cos \theta) \:,
  \\
  \beta (\rho)
  & \equiv
  \sin 2 \phi (2 \rho - \eta) (1 - \cos \theta)
  \:.
 \end{split}
 \label{equ:alpha_beta_def}
\end{equation}
When the matrix $\vect{\Sigma}$ is diagonal,
the vertical (horizontal) polarization amplitude $a_v$ ($a_h$) only
couples to the $x$ ($y$) current amplitude $j_x$ ($j_y$) and thus cross
polarization is zero.  Further, when the two diagonal elements of
 $\vect{\Sigma}$ are equal, the angular response patterns to vertical and horizontal
 polarization waves are the same and thus the differential response is zero.

\begin{figure}[tbp]
 \begin{center}
  \includegraphics[scale=0.6]{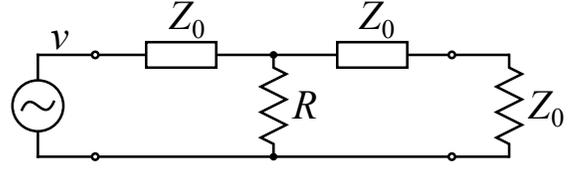}
  \caption{\label{fig:no_bs_transmission_line} 
  A transmission line of impedance $Z_0$ with a resistance $R$, modeling 
  a freestanding absorber with an on-axis incident wave.
  The incident wave with a voltage amplitude $v$ 
  induces current $j$ through the resistance,
  where $j = v \cdot 2 / (2 R + Z_0)$.
  }
 \end{center}
\end{figure}

For on-axis incidence with $\theta=0$, the relation between the incident
field and the
current is
equivalent to the case for a one-dimensional transmission
line~(Fig.~\ref{fig:no_bs_transmission_line}), as expected:
\begin{equation}
 \vect{\Sigma} \Bigr|_{\theta=0}
  =
 \left(
  \begin{array}{cc}
   2 / (2\rho_x + \eta) & 0 \\
   0 & 2 / (2\rho_y + \eta) \\
  \end{array}
 \right).
\end{equation}

\section{Analytic calculation of the response of thin absorber with
 reflective backshort termination} \label{sec:appendix_calc_with_backshort}
We adopt image theory and use the setup of
Fig.~\ref{fig:geometry_with_backshort}b.
As the mirror conjugate flips the sign of the $z$ components of vectors
and the sign of the electric field, 
the conjugate field of $\vect{E}_i$ can be written as
\begin{equation}
 \begin{split}
 \vect{E}_{\bar{i}}
  & =
  -e^{-i \vect{\bar{k}}\cdot r} \left( a_v \evb + a_h \ehb \right)
  \\
  & =
  e^{-i \vect{\bar{k}}\cdot r}
     \vect{R}_z(\phi+\pi) \left( a_\parallel \eparal + a_\perp \eperp \right)
  \:,
 \end{split}
\end{equation}
with
\begin{equation}
 \evb  \equiv
  \left(
   \begin{array}{ccc}
    1 & & \\
     & 1 & \\
     & & -1 \\
   \end{array}
  \right)
  \ev \:, \quad
 \ehb  \equiv
  \left(
   \begin{array}{ccc}
    1 & & \\
     & 1 & \\
     & & -1 \\
   \end{array}
  \right)
  \eh \:.
\end{equation}
The conjugates of the reflected and transmitted components,
$\vect{E}_{\bar{r}}$ and $\vect{E}_{\bar{t}}$, respectively,
and all the associated magnetic fields can be defined in the same way.
In particular, those for the transmitted field
 are
\begin{equation}
 \begin{split}
  \vect{E}_{\bar{t}} (\vect{r};\vect{k})
  & = e^{-i\vect{\bar{k}}\cdot\vect{r}}
  \vect{R}_z(\phi+\pi) \left( c_\parallel \eparal + c_\perp \eperp
  \right) \:,
  \\
  \vect{B}_{\bar{t}} (\vect{r};\vect{k})
  & = - \frac{1}{c} e^{-i\vect{\bar{k}}\cdot\vect{r}}
  \vect{R}_z(\phi+\pi) \left( c_\parallel \eperp - c_\perp \eparal
  \right) \:.
 \end{split}
\end{equation}
The boundary condition is defined in the same manner as
Eq.~(\ref{equ:discontinuity}) but at the surface of $z=-d$,
leading to
\begin{equation}
 \begin{split}
  \gamma  a_\parallel - \gamma^* b_\parallel
  - \gamma c_\parallel + \gamma^* c_\parallel & = 0 \:,
  \\
  \gamma  a_\perp - \gamma^* b_\perp
  - \gamma c_\perp + \gamma^* c_\perp & = 0 \:,
 \end{split}
 \label{equ:boundary1_with_bs}
\end{equation}
and
\begin{equation}
 \begin{split}
  \gamma  a_\parallel + \gamma^* b_\parallel
  - \gamma c_\parallel - \gamma^* c_\parallel & = 
  \eta (j_x \cos \phi + j_y \sin \phi) \:,
  \\
  \gamma  a_\perp + \gamma^* b_\perp
  - \gamma c_\perp - \gamma^* c_\perp & = 
  - \frac{\eta}{\cos \theta} (j_x \sin \phi - j_y \cos \phi) \:,
 \end{split}
 \label{equ:boundary2_with_bs}
\end{equation}
with $\gamma \equiv e^{ik_z d}$.
On the other hand, Ohm's law leads to
\begin{equation}
 \left(
  \begin{array}{c}
   \rho_x j_x   \\
   \rho_y j_y   \\
  \end{array}
 \right)
 =
 (\gamma - \gamma^*) \,
 \vect{R}
 \left(
  \begin{array}{c}
   c_\parallel \cos \theta \\
   c_\perp  \\
  \end{array}
 \right) \:.
 \label{equ:ohm_with_bs}
\end{equation}
Solving Eqs.~(\ref{equ:boundary1_with_bs}), (\ref{equ:boundary2_with_bs})
and (\ref{equ:ohm_with_bs}), we obtain
\begin{equation}
 \left(
  \begin{array}{c}
   j_x \\
   j_y \\
  \end{array}
 \right)
 =
 \vect{\mathcal{A}}_{\mathrm{bs}}
 \left(
  \begin{array}{c}
   a_\parallel \\
   a_\perp \\
  \end{array}
 \right) ,
  \label{equ:absorption_in_perp_parall_bs}
\end{equation}
with
\begin{equation}
  \vect{\mathcal{A}}_{\mathrm{bs}}
   \equiv
   \eta^{-1}
   \left[
    \frac{1}{\gamma - \gamma^*}
    \vect{Q}^{-1} \vect{R}^{-1} \vect{G}^{-1}
    +
    \frac{\sec \theta}{2\gamma} \vect{Q} \vect{R}^{-1}
   \right]^{-1}
   \:.
\end{equation}
Combining this with Eq.~(\ref{equ:vert_hor_to_paral_perp}), we obtain
\begin{equation}
 \left(
  \begin{array}{c}
   j_x \\
   j_y \\
  \end{array}
 \right)
 =
 \vect{\mathcal{A}}_{\mathrm{bs}} \vect{R}^{-1}
 \left(
  \begin{array}{c}
   a_v \\
   a_h \\
  \end{array}
 \right)
 \equiv
 \vect{\Sigma}_{\mathrm{bs}}
 \left(
  \begin{array}{c}
   a_v \\
   a_h \\
  \end{array}
 \right)
 , 
\end{equation}
where
\begin{equation}
 {\vect{\Sigma}_{\mathrm{bs}}}^{-1}
  =
  \frac{e^{-i k_z d}}{4 \sin k_z d \cos \theta}
 \left(
  \begin{array}{cc}
   \alpha^{\mathrm{bs}}_+ (\rho_x) & \beta^{\mathrm{bs}} (\rho_y) \\
   \beta^{\mathrm{bs}} (\rho_x) & \alpha^{\mathrm{bs}}_- (\rho_y) \\
  \end{array}
 \right) \:,
\end{equation}
with
\begin{equation}
 \begin{split}
 \alpha^{\mathrm{bs}}_\pm (\rho)
  \equiv & 
  \left[ (\rho + \eta) \sin k_z d -  i \rho \cos k_z d \right]
  \left(1 + \cos \theta \right)
  \\
  & \quad \quad \pm \delta^{\mathrm{bs}}(\rho, d, \theta) \, \cos 2\phi \:,
  \\
  \beta^{\mathrm{bs}} (\rho)
  \equiv & \delta^{\mathrm{bs}}(\rho, d, \theta) \, \sin 2\phi
   \:,
 \end{split}
\end{equation}
and
\begin{equation}
 \delta^{\mathrm{bs}}(\rho, d, \theta)
  \equiv
  \left[
  (\rho - \eta) \sin k_z d - i \rho \cos k_z d
  \right]
  \left( 1-\cos \theta \right) \:.
\end{equation}

\begin{figure}[tbp]
 \begin{center}
  \includegraphics[scale=0.6]{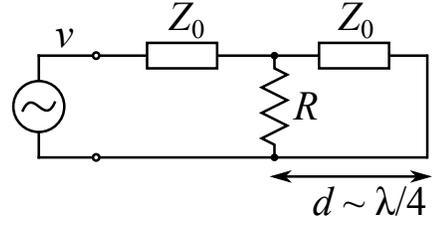}
  \caption{\label{fig:with_bs_transmission_line} 
  A transmission line of impedance $Z_0$ with a resistance $R$
  and a reflective termination, modeling 
  an absorber with reflective backshort termination
  and with an on-axis incident wave.
  The distance between the resistance and termination is $d$
  and the wavelength in the transmission line is $\lambda$.
  The incident wave with a voltage amplitude $v$ 
  induces current $j$ through the resistance,
  where
  $j = v \cdot 2 e^{i k d} \sin kd / [(R+Z_0)\sin kd - i R \cos kd]$,
  with wave number $k\equiv 2\pi/\lambda$.
  }
 \end{center}
\end{figure}

Again, for on-axis incidence, $\theta=0$, the relation between the
incident field and the
current is
equivalent to that of the one-dimensional transmission
line~(Fig.~\ref{fig:with_bs_transmission_line}), as
expected:
\begin{equation}
\vect{\Sigma}_{\mathrm{bs}} \Bigr|_{\theta=0}
 =
 \left(
  \begin{array}{cc}
   \sigma_0(\rho_x) & 0 \\
   0 & \sigma_0(\rho_y) \\
  \end{array}
 \right) \:,
\end{equation}
with
\begin{equation}
 \sigma_0(\rho)
  \equiv
  \frac{2 e^{i k d } \sin k d }{(\rho+\eta)\sin k d - i
  \rho \cos k d}
  \:.
\end{equation}

\section{Possible Systematics due to the Finite Physical Size of the
 Grids} \label{app:detector_geometry_assumption}
All the discussions in this paper are based on an assumption 
that the absorber can be modeled as a thin resistive membrane.
When the absorber is a pair of orthogonal grids,
it deviates from the ideal membrane model in the following three
ways:
1) there is a finite distance between the layers of orthogonal grids,
2) the grids have non-zero pitch and consist of wires with non-zero
cross-sections, and 3) the two orthogonal grids may couple to each other
through near-field effects.
In this Appendix, we estimate the magnitude of the systematics due to these
 aspects and discuss the conditions on the physical dimensions 
of the device required to suppress such artifacts.

\subsection{Non-zero Distance between the Two Layers}
When the absorber is a pair of resistive grids, which are sensitive to
orthogonal polarizations, there is a finite gap between the two grid
layers.  This may lead to additional beam systematics of cross
polarization and differential response.
To estimate the effect, we
consider a freestanding absorber with $\rho = \eta/2$, where there
is no systematics in the nominal configuration.

We first point out that 
the primary effect here is the cross polarization, not the differential
response.  This can be seen in Eq.~(\ref{equ:sigma_matrix_no_bs});
when $\rho_x = \eta/2$, the response of the grid running in the $x$
direction,
$G_{CO}^v$, is independent of
$\rho_y$ including the case where the grid running in the $y$ direction is
absent, or $\rho_y = \infty$.  Thus, the co-polar beam shape does not
see the effect of the other grid to first order.

\begin{figure}[htbp]
  \setlength{\unitlength}{0.15\textwidth}
  \begin{picture}(1.6,1.656)
   \put(0,0.1){\includegraphics[width=1.6\unitlength]{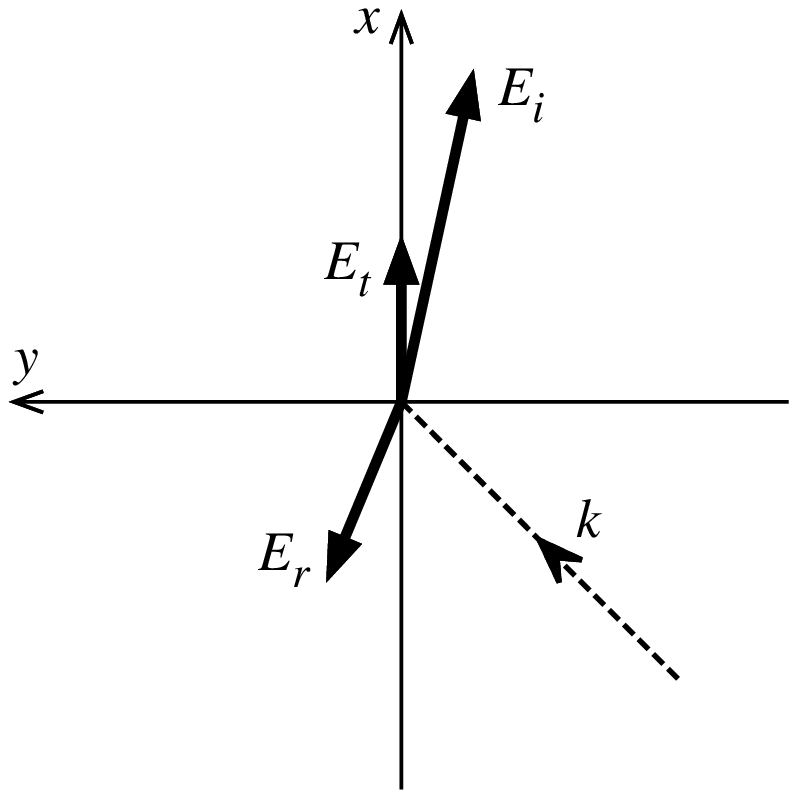}}
   \put(0,1.65){(a)}
  \end{picture}
  \hspace{0.03\textwidth}
  \setlength{\unitlength}{0.15\textwidth}
  \begin{picture}(1.0,1.656101)
    \put(0,0){\includegraphics[width=\unitlength]{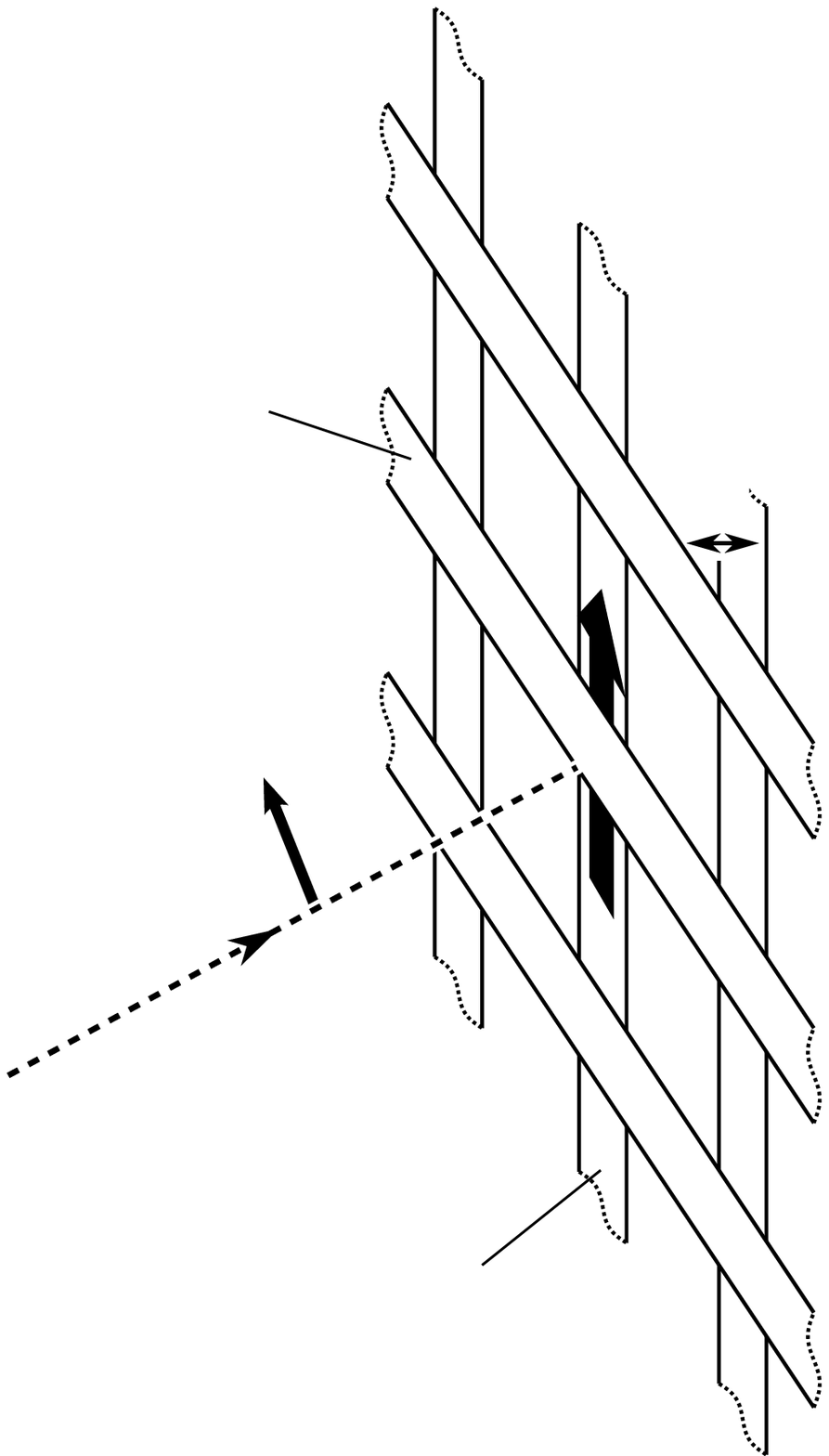}}
    \put(0.85,1.14){\large $\varepsilon$}
    \put(0.16,0.8){$\vect{E}_i$}
    \put(-0.00100558,1.65){(b)}
    \put(0.76,0.87){$\vect{J}$}
    \put(0.38,0.14){$x$ grid}
    \put(0.11,1.3){$y$ grid}
    \put(0.25,0.48){$\vect{k}$}
  \end{picture}
 \caption{ \label{fig:vect_e_finite_distance} 
 (a) The electric field of a {\it vertically} ($x$) polarized incident wave
 with $(\theta, \phi) = (50^\circ, 45^\circ)$
 projected on the plane of the absorber (the $x$-$y$ plane).
 (b) A schematic figure showing the case with a gap $\varepsilon$
 in the $z$ direction
 between the planes of $x$ and $y$ grids,
 with the {\it vertically} polarized incident wave.
 }
\end{figure}

Cross-polarization, on the other hand, can arise from the finite
gap.  Consider the electric fields of incident,
reflected, and transmitted waves projected on the absorber surface
plane ($x$-$y$ plane), $\vect{E}_i \bigr|_s$, $\vect{E}_r \bigr|_s$, and
$\vect{E}_t \bigr|_s$, respectively, for vertical polarization incident.
Figure~\ref{fig:vect_e_finite_distance}a shows an example with an
incident angle of $\phi=45^\circ$,
where the cross polarization is maximum, and $\theta=50^\circ$.  The
boundary condition
guarantees
$\vect{E}_i \bigr|_s + \vect{E}_r \bigr|_s = \vect{E}_t \bigr|_s$.
Zero cross-polarization of the nominal configuration corresponds to
 a vanishing $y$ component of the field:
$E_{i, y} + E_{r, y} = E_{t, y} = 0$.  However, this cancellation of the $y$
components of the incident and reflected fields is only exact on the
surface of the grid running in the $x$ direction, which we define as $z=0$.
When the other grid
running in $y$ direction is placed at $z = -\varepsilon$
(Fig.~\ref{fig:vect_e_finite_distance}b), the $y$ grid
feels a residual electric field of order
\begin{equation}
 (1-\cos \theta) \sin k_z \varepsilon  
  \simeq 2 \pi \frac{\varepsilon}{\lambda}
  \cos \theta (1-\cos \theta) \leq \frac{\pi \varepsilon}{2\lambda} \:.
\end{equation}
Thus, the cross-polar level is
\begin{equation}
 XP \sim \left( \pi \varepsilon / 2 \lambda \right)^2 \:.
\end{equation}

\subsection{Finite Pitch and Cross Section of the Grid}
Compared to an ideal resistive sheet,
a grid of resistive wires has non-zero spacing between the grid wires and
non-zero cross section of each wire.  These can cause deviations from the ideal
current sheet when the physical dimensions are not small compared
to the wavelength $\lambda$.
The requirements for the 
smallness are discussed in detail in literature (see, e.g.,
Refs.~\cite{1962ITMTT..10..191L,2012ApOpt..51..197C}
and references therein) in the
context of grids made of conductive wires.  
Estimates for the grid of resistive strips do not significantly
deviate from conductive wires when the resistivity is
similar to or less than
the impedance of free space,
$\rho \lesssim \eta$~\cite{1987ITAP...35.1492G,Zinenko1998},
which is the parameter space of interest here.
In the regime
$\lambda \gg g > 4\pi a$, where $g$ is the spacing between the
wires and $a$ is
the radius of each wire of circular cross-section,
the first order estimates of the cross-polar level and the differential
response are (see, e.g., Ref.~\cite{1962ITMTT..10..191L})
\begin{equation}
 XP \sim \left( \frac{2 \pi^2 a^2}{\lambda g} \right)^2 \:,
  \quad
  DR \sim \left( \frac{2 g}{\lambda} \ln \frac{g}{2\pi a} \right)^2 \:.
  \label{equ:xp_dr_of_finite_pitch_and_cross_section}
\end{equation}
An estimate for a grid made of thin strips of width $w$ can be
obtained by substituting $w = 2a$, to first order.
In the regime $\lambda \gg 4\pi a \gtrsim g$, the
approximation of the term involving $\ln g/ 2\pi a$ becomes
invalid as pointed out in Ref.~\cite{2012ApOpt..51..197C}.  However, the
numerical analysis in the reference shows that the term is
still a monotonically decreasing function of $a/g$.  Thus, one can
still put a rough upper limit on the systematics based on
Eq.~(\ref{equ:xp_dr_of_finite_pitch_and_cross_section}) 
and the monotonic dependence on $a/g$ even in this regime.
In the implementation of Ref.~\cite{2011JCAP...07..025K}, for example, 
the parameters are
$a \simeq 1\,\mathrm{\mu m}$ and $g \simeq 30\,\mathrm{\mu m}$,
and satisfy $\lambda \gg g > 4\pi a$.
The systematics in this case are
$XP \sim \left( 1 / \lambda [\mathrm{\mu m}] \right)^2$
and $DR \sim \left( 10/\lambda [\mathrm{\mu m}] \right)^2$, negligibly
 small for millimeter and submillimeter wavelengths.

\subsection{Near-field Coupling}
All the discussions above assume only the
far-field effects, or the radiating modes,
and we ignored the near-field effects due to the evanescent modes.
Here we show that the near-field
coupling between the orthogonal grids
is small when the distance between the grids, $\varepsilon$,
is similar to or larger than the grid spacing, $g$.
Owing to the periodic symmetry of the system,
the scattered field including the evanescent modes
can be expanded in terms of a Floquet
series~\cite{Zinenko1998,Matsushima2000}.
The phase component of the Floquet series is
\begin{equation}
 e^{-i k_{z,mn} \, \left|\Delta z\right|}
  e^{- i \left( k_{x,m} \, x + k_{y,n} \, y \right)} \:,
\end{equation}
with
\begin{equation}
\begin{split}
 k_{x,m} & \equiv k_x + 2 \pi m / g \:, \quad
 k_{y,n} \equiv k_y + 2 \pi n / g \:, 
 \\
 k_{z,mn} & \equiv \sqrt{ k^2 - {k_{x,m}}^2 - {k_{y,n}}^2 }
 \quad (\mathrm{Im} [ k_{z,mn} ] \leq 0) \:,
\end{split}
\end{equation}
where $\Delta z$ is the distance from the scattering grid,
and $m$ and $n$ are the integer numbers corresponding to the series
indices.
Our interest here is the non-radiating modes, or the modes with
$m \neq 0$ or $n \neq 0$.
When the grid spacing is small compared to the wavelength
($g \ll \lambda$),
the non-radiating mode has
\begin{equation}
 k_{z,mn} \simeq
  - i \frac{2\pi}{g}  \sqrt{m^2 + n^2} \:,
\end{equation}
and the field strength decays as
$e^{- 2\pi \left| \Delta z \right| / g}$
or faster.
Thus, the near-field coupling between the paired grids
 is suppressed by a factor of
 $e^{-2\pi \varepsilon/g}$.
 This factor is small 
  when $\varepsilon$ is similar to or larger than the grid
 pitch ($\varepsilon \gtrsim g$).
Therefore, the near-field effect can safely be ignored in this regime
of $\varepsilon \gtrsim g$, with $g \ll \lambda$.

\section{Relation among Resistive Sheet, Conic Reflector, and Parallel
 Current on Their Surface \label{app:koffman_relation}
}
Koffman showed the condition where the induced current on a conic
 reflector illuminated by a feed 
flows in parallel~\cite{1966ITAP...14...37K}.
The condition is met when the magnetic field $\vect{H}$ at the reflector surface satisfies
\begin{equation}
 \frac{H_\parallel}{H_\perp} = \tan \phi \cdot
  \frac{\kappa + \cos
  \theta}{1 + \kappa \cos \theta}
  \:,
 \label{equ:koffman_gemetry_hh}
\end{equation}
where the angle $(\theta,\phi)$ specifies a position on the reflector 
with the focus as the origin of the coordinates, $H_\parallel$ and $H_\perp$
are the magnetic field
along $\eq$ and $\ep$, respectively, and $\kappa$ is the eccentricity
specifying the geometry of the conic reflector: $\kappa=0$,
$0<\kappa<1$, $\kappa=1$, $1<\kappa<\infty$, and $\kappa=\infty$
correspond to sphere, ellipsoid, paraboloid, hyperboloid, and plane,
respectively.
Koffman pointed out that a feed radiation pattern satisfies
Eq.~(\ref{equ:koffman_gemetry_hh}) when it can be expressed as a
superposition of electric- and magnetic-dipole radiations, and relative
strengths between them coincide with the eccentricity $\kappa$.
Note that the object is at infinitely far for astronomical applications
and thus a paraboloid reflector or an equivalent system~\cite{Mizuguchi,Dragone}
is usually employed.  This is why we adopted Ludwig's third definition,
in which a pure polarization pattern satisfies Eq.~(\ref{equ:koffman_gemetry_hh})
with $\kappa = 1$.

We can see two connections between this symmetry explored by Koffman and our
result for a freestanding resistive absorber sheet.
One of them can be seen by replacing the feed in Koffman's setup by a
resistive sheet.  
In the replacement, we also replace the radiation field pattern from a
pure-polarization feed by the absorption field pattern that induces
current in only $x$, i.e., $j_y = 0$.  
Equations~(\ref{equ:perp_paral_theta_phi}), (\ref{equ:wave_incident}), and
(\ref{equ:vert_hor_to_paral_perp}) lead to
$\eta (H_\perp,-H_\parallel)^T = (a_\parallel,a_\perp)^T = \vect{R}(a_v,a_h)^T$.
Thus, Eqs.~(\ref{equ:j_vs_sigma_vs_ampl}), (\ref{equ:sigma_def}), 
(\ref{equ:alpha_beta_def}),
and $j_y=0$ lead to
\begin{equation}
 \frac{H_\parallel}{H_\perp} = \tan \phi \cdot
  \frac{\frac{\eta}{2\rho} + \cos
  \theta}{1 + \frac{\eta}{2\rho} \cos \theta}
  \label{equ:sheet_hh_ratio}
\end{equation}
for the reception pattern of the thin resistive absorber.
Equations~(\ref{equ:koffman_gemetry_hh}) and (\ref{equ:sheet_hh_ratio})
have equivalent form, due to the following parallel between Koffman's
and our results: $\kappa \rightarrow 0$ ($\rho \rightarrow \infty$)
corresponds to magnetic dipole (magnetic conductor sheet),
while $\kappa \rightarrow \infty$ ($\rho \rightarrow 0$)
corresponds to electric dipole (electric conductor sheet);
the electric and magnetic contributions balances when
$\kappa = 1$ ($\rho = \eta/2$).
See also the
discussion at the end of Sec.~\ref{sec:no_backshort}. 

Another connection between Koffman's and our results can be seen by
replacing the conic reflector in Koffman's setup with a freestanding
thin resistive absorber.  Namely, one illuminates the resistive absorber
by a superimposed electric- and magnetic-dipole radiations
with their relative strength of $\kappa$.  According to the derivation
above, the induced current on the absorber flows in parallel when the
sheet resistivity satisfies $\kappa = \eta / 2\rho$.
Thus, instead of the eccentricity of the conic reflector,
one can use the resistivity of the absorbing sheet 
as the parameter tuned to match with an arbitrary superposition of the
dipoles.

\end{document}